# Optimal internal boundary control of lane-free automated vehicle traffic

Milad Malekzadeh[1], Ioannis Papamichail[1], Markos Papageorgiou[1], Klaus Bogenberger[2]

[1]Dynamic Systems and Simulation Laboratory, Technical University of Crete, Chania 73100, Greece

[2]Technical University of Munich, Chair of Traffic Engineering and Control, Munich, Germany

**Abstract**

A recently proposed paradigm for vehicular traffic in the era of CAV (connected and automated vehicles), called TrafficFluid, involves lane-free vehicle movement. Lane-free traffic implies that incremental road widening (narrowing) leads to corresponding incremental increase (decrease) of capacity; and this opens the way for consideration of real-time internal boundary control on highways and arterials, in order to flexibly share the total (both directions) road width and capacity among the two directions in dependence of the bi-directional demand and traffic conditions, so as to maximize the total (two directions) flow efficiency. The problem is formulated as a convex QP (Quadratic Programming) problem that may be solved efficiently, and representative case studies shed light on and demonstrate the features, capabilities and potential of the novel control action.

## 1. Introduction

Vehicular traffic is crucial for the transport of persons and goods, but traffic congestion, which appears on a daily basis, particularly in and around metropolitan areas, around the world has been an increasingly serious problem that calls for drastic solutions. Traffic congestion causes substantial delays, excessive environmental pollution and reduced traffic safety. Conventional traffic management measures are valuable (Papageorgiou et al., 2003; Papageorgiou et al., 2007; Kurzhanskiy and Varaiya, 2010), but not always sufficient to tackle the heavily congested traffic conditions, which must be addressed in a more comprehensive way that exploits gradually emerging and future ground-breaking new capabilities of vehicles and the infrastructure.

During the last decade, there has been an enormous effort by the industry and by numerous research institutions to develop and deploy a variety of vehicle automation and communication systems that are revolutionizing the vehicle capabilities. Vehicle automation ranges from different kinds of driver support to highly or fully automated driving; and vehicle communication enables V2V (vehicle-to-vehicle) and V2I (vehicle-to-infrastructure) communication that may support various potential applications. Many automotive and information-technology companies, as well as research institutions, have been developing and testing in real traffic conditions high-automation or virtually driverless autonomous vehicles that monitor their environment and make sensible driving decisions based on appropriate decision and control methods (Ardelt et al., 2012; Aeberhard et al., 2015; Kamal et al., 2016; Makantasis and Papageorgiou, 2018).

A recent paper (Papageorgiou et al., 2021) launched the TrafficFluid concept, which is a novel paradigm for vehicular traffic, applicable at high levels of vehicle automation and communication (SAE levels 4 or 5) and high penetration rates that are expected to prevail in the not-too-far future. The TrafficFluid concept is based on the following two combined principles: (1) *Lane-free traffic*, whereby vehicles are not bound to fixed traffic lanes, as in conventional traffic, but may drive anywhere on the 2-D surface of the road; (2) *Vehicle nudging,* whereby vehicles communicate their presence to other vehicles in front of them (or are sensed by them), and this may exert a “nudging” effect on the vehicles in front, i.e. vehicles in front may, under appropriate circumstances, experience (apply) a pushing influence. Several advantages and challenges related to this novel traffic paradigm are discussed by Papageorgiou et al. (2021) (see also (Papageorgiou et al. 2019) for more modelling details).

This paper exploits the lane-free property of TrafficFluid, i.e. the possibility for vehicles to drive on the 2-D road surface without being bound to lanes. As demonstrated in a small experiment by Papageorgiou et al. (2021) and is also intuitively sensible, lane-free traffic implies that the traffic capacity may exhibit incremental (increasing or decreasing) changes in response to corresponding incremental (widening or narrowing) changes of the road width. This is in contrast to lane-based roads and traffic, where capacity changes may only occur if the road width is changed by lane “quanta”.

To illustrate this feature of lane-free connected and automated vehicle (CAV) traffic, the snapshot of a lane-free microscopic simulation video using an ad-hoc vehicle movement strategy (see (Papageorgiou et al., 2019) for modelling details) in Fig. 1 illustrates that, on a highway with constant road width, vehicles, driven by their smooth (no lane changes) 2-D movement strategy, spread on the available 2-D road surface without necessarily forming lanes, except, at times, along the road boundaries, so as to maximize the infrastructure utilization. If the road would be widened by some amount, vehicles would immediately move laterally to cover the free space. This would increase the average inter-vehicle spacing, thus allowing for higher vehicle speeds and hence higher flow and capacity.

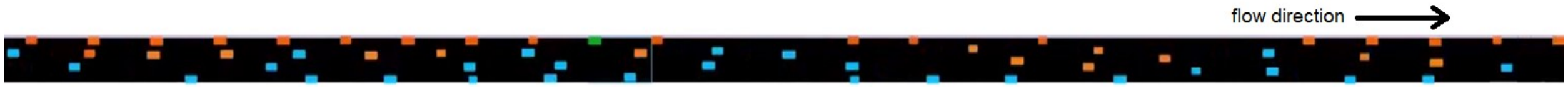

*Figure 1. Snapshot of microscopic simulation with lane-free CAV traffic*

Consider a road or highway with two opposite traffic directions, where CAV are driving. The total carriageway width, and hence the total capacity (for both directions), could be shared among the two directions in a flexible way, according to the prevailing demand, so as to maximize the infrastructure exploitation. Flexible width and capacity sharing may be achieved via virtual moving of the internal boundary, which separates the two traffic directions, and corresponding communication to the CAV to respect the changed internal boundary. Such internal boundary shift includes shifting of possible safety margins among the two directions. This way, the carriageway's width portion (and corresponding total capacity share) assigned to each traffic direction can be changed in space and time (subject to constraints) according to an appropriate control strategy, as illustrated in Fig. 2, so as to maximize the total traffic efficiency in both directions.

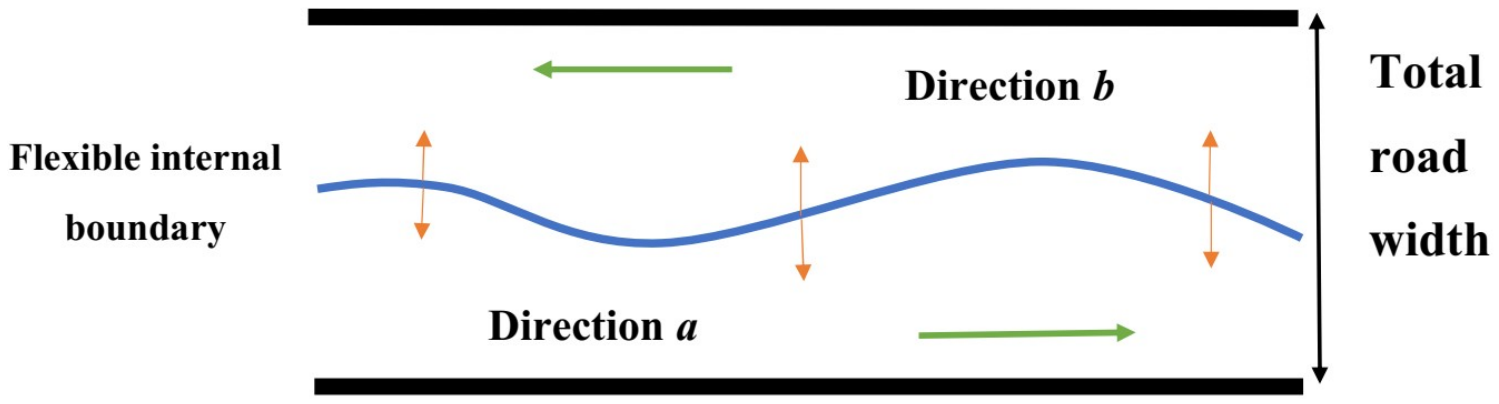

*Figure 2. Space-time flexible internal road boundary.*

The idea of sharing the total road capacity among the two traffic directions is not new and has been occasionally employed for conventional lane-based traffic in many countries, typically offline or manually (Wolshon and Lambert, 2004). The measure is known as tidal flow or reversible lane control, and its main principle is to adapt the total available cross-road supply to the bi-directional demand. Its most basic form is the steady allocation of one (or more) lanes of one direction to the opposite direction for a period of time (ranging from few hours to many days) in the aim of addressing abnormal traffic supply or demand. This happens indeed often at work zones, in order to compensate for the capacity loss in one direction due to road works; or in cases of big events (sport events, concerts, holiday departure or return, evacuation etc.) due to excessive demand in one traffic direction, while demand in the opposite direction is low. More advanced reversible lane systems may operate in real time, see (Frejo et al., 2015; Ma, et al., 2018; Ampountolas et al., 2020), to balance delays on both sides of a known bottleneck (e.g. bridge or tunnel) by assigning a lane to the each of the two directions in alternation in response to the prevailing traffic conditions. To this end, optimal control or feedback control algorithms of various types are proposed by Frejo et al. (2015), Ma et al. (2018) and Ampountolas et al. (2020).

Reversible lanes have also been considered in connection with lane-based CAV driving. To start with, Hausknecht et al. (2011) formulate an integer linear problem and find the lane configuration that is able to serve the maximum amount of traffic a network can handle satisfying constant demand for each origin-destination pair. They also formulate a bi-level programming problem, whereby the upper level includes the allocation of capacity to each of the links, while the lower level addresses User Equilibrium (UE) conditions; and employ genetic algorithms for the solution. Duell et al. (2015; 2016) use the system optimal dynamic traffic assignment models formulated by Ziliaskopoulos (2000) for a single destination and by Li et al. (2003) for more general networks, using the Cell Transmission Model (CTM) by Daganzo (1994). Lanes are reflected with integer variables, and the problem is formulated as a mixed integer linear programming (MILP) problem that has, however, high (exponential) complexity due to the many integers variables involved. Levin and Boyles (2016) use this model for a single link and utilize stochastic demand as a Markov decision process. The MILP problem is solved using a heuristic and is incorporated within a UE routing problem. Finally, Chu et al. (2019) address the reservation-based routing and scheduling of CAV, considering dynamically reversible lanes, as an integer linear program.

Lane-based tidal flow control systems for conventional (manually driven vehicle) traffic may be very useful in certain situations, but they face a number of difficulties that limit their widespread use:

- The resolution of infrastructure sharing among the two traffic directions cannot be higher than one lane, which may not be sufficiently fine-grained for many traffic situations.
- A reversible lane must extend over sufficient length (minimum of few kilometers) to mitigate counter-problems due to frequent merging or diverging of traffic streams.
- Whenever a reversible lane switch to the opposite direction is decided, a time-delay (corresponding to the travel time on the reversible lane) must be respected, before actually opening the lane to the opposite direction, so as to allow for the evacuation of the lane and avoid simultaneous opposite-direction movements. For the duration of this delay, the reversible lane is under-utilized, and this side-effect is an overhead that reduces the overall benefit of the measure. To mitigate the impact of this side-effect, the frequency of reversible lane switches must be limited (Frejo et al., 2015).

Due to these limitations, reversible lane control has not evolved as a major traffic management measure, even less so in real time. Even for future CAV traffic, however, some of the mentioned difficulties would persist in lane-based conditions, notably the low capacity sharing resolution, the merging nuisance and, last but not least, the complex (integer-based) nature of the corresponding optimization problems that hinder real-time feasibility.

In contrast, in a lane-free CAV traffic environment, there is a clear prospect for the mentioned difficulties to be largely mitigated. Specifically:

- The resolution of infrastructure sharing among the two directions can be high, still leading to corresponding intended capacity changes for the two opposite traffic streams.
- The potentially smooth (no lane changes) driving of CAV in a lane-free road surface would allow for the internal boundary to be a smooth space-function, as illustrated in Fig. 2. This function may be softly changing in real time in response to the prevailing traffic conditions.
- Due to moderate changes of the internal boundary over time and space and the lack of physical boundary, CAV may respect relatively promptly the changed boundary; hence the aforementioned safety-induced time-delay, required to avoid opposite movements on the same road surface, may be small.
- As practiced in this paper, the resulting optimization problems include only real-valued variables (no integers are necessary) and may therefore be solved very efficiently, so as to be readily real-time feasible.

Thanks to these characteristics, real-time internal boundary control for lane-free CAV traffic has the potential to be broadly applicable to the high number of arterial or highway infrastructures that feature unbalanced demands during the day in the two traffic directions, so as to strongly mitigate or even avoid congestion. Even for infrastructures experiencing strong demand in both directions during the peak periods, real-time internal boundary control may intensify the road utilization and lead to sensible improvements. In fact, the possibility to control the internal boundary in real time may prove to be one of the major advantages of the TrafficFluid concept.

This paper proposes a macroscopic model-based optimization scheme to elaborate on and demonstrate the characteristics of internal boundary control. The well-known CTM (Daganzo, 1994) is employed to this end, leading to a convex Quadratic Programming (QP) problem. Carefully designed simulation scenarios highlight some very interesting and intriguing implications of this innovative control measure. Section 2 presents the general problem statement, followed by the CTM presentation, the transformation of CTM equations to linear equality and inequality constraints, and the proposed quadratic cost function. Two considered case studies are introduced in Section 3, along with the obtained results. Conclusions and on-going work are summarized in Section 4.

## 2. Optimal internal boundary control scheme

### 2.1. General modelling and problem statement

The low lane discipline and vehicle size diversity encountered in several developing countries motivated, in the last few years, some microscopic modelling works, which proposed, using various approaches, models for heterogeneous and lane-less traffic; similar developments were triggered by the recent interest in shared spaces, which are used by pedestrians and vehicles of various types; see e.g. (Manjunatha et al., 2013; Rudloff et al., 2013); see also (Kanagaraj and Treiber, 2018; Mulla et al., 2018), where such models are proposed, validated with real traffic data and analyzed with respect to stability and other properties. Clearly, these modelling works attempt to describe the driving behavior of real vehicles and drivers. In the case of lane-free CAV traffic, as proposed in the TrafficFluid concept, we need to design (rather than model) opportune movement strategies for safe and efficient traffic flow. Such an ad-hoc CAV movement strategy was reported in (Papageorgiou et al., 2019), but more systematic developments in this direction are in progress.

Lane-free traffic is not expected to give rise to structural changes of existing macroscopic models. It is reasonable to assume, as also supported by results in (Bhavathrathan and Mallikarjuna, 2012; Asaithambi et al., 2016; Munigety et al., 2016; Papageorgiou et al., 2019), that notions and concepts like the conservation equation, the fundamental diagram, as well as moving traffic waves will continue to characterize macroscopic traffic flow modelling in the case of lane-free CAV traffic. By the same token, specific physical traffic parameters, such as free speed, critical density, flow capacity, jam density, are also relevant for lane-free traffic, but may of course take different values than in lane-based traffic. However, the exact values that these parameters will take in lane-free CAV traffic are of minor importance for the presented demonstration of a novel control measure.

In the present context, it is crucial to elaborate on the impact of internal boundary control on the respective Fundamental Diagrams (FDs) of the two opposite traffic directions, which we will call directions $a$ and $b$, respectively (Fig. 2). Let us assume that directions $a$ and $b$ are assigned respective road widths (in m) $w^a = \varepsilon \cdot w$ and $w^b = (1-\varepsilon) \cdot w$, where $0 \le \varepsilon \le 1$ is the sharing factor and $w$ is the total road width (both directions). Let $Q(\rho)$, where $\rho$ is the traffic density in veh/km, be the total FD (both directions) of a highway section, which would prevail if the whole carriageway would be assigned to only one of the two opposite traffic directions (i.e. for $\varepsilon$ equal to 0 or 1), with total critical density $\rho_{cr}$, total capacity $q_{cap}$ and total jam density $\rho_{\max}$. Let us consider the case of partial road sharing, i.e. $\varepsilon_{\min} \le \varepsilon \le \varepsilon_{\max}$, where $\varepsilon_{\min}, \varepsilon_{\max} \in (0,1)$ are appropriate bounds to be specified later. We want to derive the corresponding FDs and parameter values for the two directions $a$ and $b$. For easier understanding, let us assume for a moment that $w$ is the total number of lanes and $\varepsilon \cdot w$, $(1-\varepsilon) \cdot w$ are the respective integer numbers of lanes assigned to the two directions. Then, $Q(\rho^1 \cdot w)/w$ is the FD per lane, where $\rho^1$ is the density per lane, and, analogously, the FDs for the two directions, which are functions of $\varepsilon$, are given by

$$Q^a(\rho^a, \varepsilon) = \varepsilon \cdot Q(\rho^a / \varepsilon),\ Q^b(\rho^b, \varepsilon) = (1-\varepsilon) \cdot Q(\rho^b / (1-\varepsilon)), \tag{1}$$

where $\rho^a$ and $\rho^b$ (in veh/km) are the respective densities of the two directions. Clearly, (1) applies in the exact same way if we return to the original meaning of $\varepsilon$ as the real-valued sharing factor.

We now wish to derive the shared critical density $\rho_{cr}^a(\varepsilon)$, capacity $q_{cap}^a(\varepsilon)$, and jam density $\rho_{\max}^a(\varepsilon)$ for direction $a$ as functions of the sharing factor $\varepsilon$. To this end, we need the derivative $\frac{\partial Q^a(\rho^a,\varepsilon)}{\partial \rho} = Q^{a\prime}(\rho^a,\varepsilon)$, for which, using (1), we obtain $Q^{a\prime}(\rho^a,\varepsilon) = Q'(\rho^a/\varepsilon)$. For the critical density, this derivative equals zero, i.e. we have $Q'(\rho_{cr}^a/\varepsilon)=0$, but we also have $Q'(\rho_{cr})=0$, hence we deduce that $\rho_{cr}^a(\varepsilon)=\varepsilon\cdot\rho_{cr}$. For the capacity, we have $q_{cap}^a(\varepsilon)=Q^a(\rho_{cr}^a,\varepsilon)=\varepsilon\cdot Q(\rho_{cr})=\varepsilon\cdot q_{cap}$. Finally, we have $Q(\rho_{\max})=Q^a(\varepsilon.\rho_{\max},\varepsilon)=0$, and hence $\rho_{\max}^a(\varepsilon)=\varepsilon\cdot\rho_{\max}$. Similarly, we have $\rho_{cr}^b(\varepsilon)=(1-\varepsilon)\cdot\rho_{cr}$, $q_{cap}^b(\varepsilon)=(1-\varepsilon)\cdot q_{cap}$ and $\rho_{\max}^b(\varepsilon)=(1-\varepsilon)\cdot\rho_{\max}$ for direction $b$.

In short, the sharing factor $\varepsilon$ scales density (veh/km) and flow (veh/h), leaving unaffected the speed (km/h). The derived equations for the FD parameters can be easily shown to hold also for FDs that are not differentiable at $\rho_{cr}$, such as the triangular FD, as we will see later, see also (Ampountolas et al., 2020).

The above derivations rely on two main assumptions. The first assumption, which is the most crucial one, is that any incremental widening (narrowing) of the road width entails a corresponding incremental increase (decrease) of capacity; or, in other words, that the capacity is a monotonic continuous function of the sharing factor, i.e. we have $q_{cap}^a(\varepsilon)$ and $q_{cap}^b(\varepsilon)$ as continuous functions. Indeed, the highway may hold vehicles of different dimensions and speeds. These vehicles occupy, in a lane-free structure, the road surface according to their movement strategies, which generate a "spread" of lateral vehicle positions, as illustrated earlier (Fig. 1). Thus, every incremental widening of the road increases the average inter-vehicle spacing and offers possibilities for higher speed, and hence higher flow and capacity. This is evidenced in FDs collected from microscopic simulation experiments and reported in (Papageorgiou et al., 2019). It should be noted that the capacity change, entailed by road width change, may be stochastic in practice and even subject to occasional random discontinuities due to the virtually unpredictable exact outcome of 2-D vehicle movement on the 2-D road surface. However, using continuous functions $q^a_{cap}(\varepsilon)$ and $q^b_{cap}(\varepsilon)$ is deemed a reasonable modelling idealization, like many others in traffic flow and other kinds of modelling.

The second, less crucial assumption is that the functions $q^a_{cap}(\varepsilon)$ and $q^b_{cap}(\varepsilon)$ are linear and symmetric, i.e. that have $q^a_{cap}(\varepsilon) = q^b_{cap}(1-\varepsilon)$. Linearity seems reasonable, and FDs reported in (Papageorgiou et al., 2019) verify it, as capacity was found to increase roughly proportionally to the sharing factor. Note, however, that the relations have $q^a_{cap}(\varepsilon)$ and $q^b_{cap}(\varepsilon)$ depend on the 2-D CAV movement strategy employed, and functional forms other than linear cannot be excluded. On the other hand, symmetry is present if the total FD $Q(\rho)$ is the same in either direction. Due to road grade or other reasons, we may sometimes have asymmetric cases with the total FD being different, i.e. $Q^A(\rho)$ in direction $a$ and $Q^B(\rho)$ in direction $b$. Cases involving nonlinear or non-symmetric functions $q^a_{cap}(\varepsilon)$ and $q^b_{cap}(\varepsilon)$ call for a slightly generalized treatment along the same lines, but are not considered here for simplicity.

A general macroscopic dynamic traffic model for internal boundary control may be expressed in discrete-time state-space form as follows:

$$\mathbf{x}^a(k+1) = \mathrm{f}^a[\mathbf{x}^a(k), \boldsymbol{\varepsilon}(k_c), \mathbf{d}^a(k)], \tag{2}$$

$$\mathbf{x}^b(k+1) = \mathrm{f}^b[\mathbf{x}^b(k), \boldsymbol{\varepsilon}(k_c), \mathbf{d}^b(k)], \tag{3}$$

where $\mathbf{x}^a$ and $\mathbf{x}^b$ are the state vectors for traffic directions $a$ and $b$, respectively, comprising section-based traffic densities and, in case of second-order models, also mean speeds; the model time step is typically 5 – 10 s for section lengths of some 500 m, and $k = 0,1,\ldots$ is the corresponding discrete time index; $\mathbf{d}^a$ and $\mathbf{d}^b$ are vectors of external variables in the respective traffic directions $a$ and $b$ (upstream mainstream demand, on-ramp flows etc.); and $\boldsymbol{\varepsilon}$ is the vector of the sharing factors (control variables), one per section. The control time step $T_c$ does not need to be equal to the model time step $T$, but is assumed to be a multiple of $T$, in which case, the control time index is given by $k_c = \lfloor kT/T_c \rfloor$, where $\lfloor . \rfloor$ is the integer part notation. It is noted that the notation $\varepsilon(k_c)$ indicates that the specific sharing factor is applied for the duration of the control time interval $[k_c \cdot T_c,(k_c+1)\cdot T_c)$. Some constraints regarding the control input will be detailed in the next section. The control inputs $\varepsilon_i$, one per section $i$, will need to be spatially smoothed prior to their actual application to the road width. This may be done via appropriate spline interpolation, the details of which will be elaborated in future work, which will test the presented concept via microscopic simulation, where safety as well as convenience implications will be carefully addressed.

Once a dynamic model is selected, the general state-space equations (2), (3) may be used to evaluate different control strategies for specific scenarios. Beyond the road's traffic characteristics, a scenario over a time horizon $K$ is defined by the initial state $\mathbf{x}^a(0)$, $\mathbf{x}^b(0)$ and the trajectories of the external variables $\mathbf{d}^a(k)$, $\mathbf{d}^b(k)$, $k=0,1,.....,K-1$. The control variables determine the share of the overall road width (and capacity) among the two opposite directions as a function of space (sections) and time (control time steps). Evaluation of the quality of a control strategy calls for the specification of an objective criterion to be minimized, as will be detailed in Section 2.2.3.

### 2.2. CTM-based optimal control problem

#### 2.2.1. CTM relevance

While other control design approaches are possible and indeed in preparation, we have opted, as a first approach, to formulate this novel traffic control action as a model-based optimal control problem. Optimization methods, when properly applied, are known to deliver solutions that may reveal novel control aspects (Papageorgiou, 1997), something that is particularly interesting in the case of still unexplored control measures, like internal boundary control.

Various dynamic traffic flow models have been employed for optimal control problems, among which a simple but realistic possibility is CTM (Daganzo, 1994); see (Ziliaskopoulos, 2000; Gomes and Horowitz, 2006; Roncoli et al., 2015) for CTM-based optimal control formulations (among many others). CTM is a first-order model deriving from the LWR model (Lighthill and Whitham, 1955; Richards, 1955), which attains a space-time discretized form by application of the Godunov numerical scheme. CTM employs a triangular FD, and its main advantage, when used within an optimal control setting, is that it may lead to a convex, hence globally optimizable, linear or quadratic programming problem, which can be solved numerically using very efficient available codes. The reason behind this property is that the nonlinearities that every traffic flow model must necessarily feature to realistically reflect the traffic flow dynamics, have, in CTM, a piecewise linear form that is amenable to linear constraints for the optimization problem; and, hence, to a convex admissible region. Thus, even large-scale traffic control problems may be solved very efficiently computationally, something that allows for real-time model predictive control (MPC). MPC is widely applied in many control application areas; and is employed, in particular, in hundreds of published works on traffic control (Burger et al., 2013). MPC's feedback structure allows for correction of inaccuracies of the employed model and demand predictions over time. MPC is not directly considered in this work, as our primary focus is to develop its kernel optimal control problem for internal boundary control.

The LWR model, and hence also CTM, are strictly based on the following three principles:

1. The conservation-of-vehicles, which is a universal physical law applying to any kind of fluid and certainly also to lane-free CAV traffic.
2. The FD with a characteristic inverse-U shape (triangular for CTM). As explicitly evidenced in (Papageorgiou et al., 2019), such a flow-density relation applies indeed also to lane-free CAV traffic.
3. The transport equation $q = \rho v$, which is also a universal physical law applying to any kind of fluid and certainly also to lane-free CAV traffic.

Given these facts, CTM appears to offer a good basis for our purposes.

In the following, we will use CTM, appropriately adjusted to incorporate the internal boundary control action, so as to cast the control problem in the form of a convex Quadratic Programming (QP) problem.

### 2.2.2. CTM equations

Consider a highway stretch holding two opposite traffic directions $a$ (from left to right) and $b$ (from right to left). The stretch is subdivided in $n$ road sections, each some 500 m in length. As explained in the previous section, the total road width, which is assumed constant over all sections for simplicity, can be flexibly shared among the two directions in real time. As the sharing may be different for every section, we have corresponding sharing factors $\varepsilon_i,\ i=1,2,.....,n$; and (1) applies to each section. As a consequence, the total section capacity, as well as the critical density and jam density, are shared among traffic directions $a$ and $b$ according to

$$\begin{aligned}
&q^{a}_{i,cap}(\varepsilon_i)=\varepsilon_i\cdot q_{cap},\ q^{b}_{i,cap}(\varepsilon_i)=(1-\varepsilon_i)\cdot q_{cap},\\
&\rho^{a}_{i,cr}(\varepsilon_i)=\varepsilon_i\cdot \rho_{cr},\ \rho^{b}_{i,cr}(\varepsilon_i)=(1-\varepsilon_i)\cdot \rho_{cr},\\
&\rho^{a}_{i,\max}(\varepsilon_i)=\varepsilon_i\cdot \rho_{\max},\ \rho^{b}_{i,\max}(\varepsilon_i)=(1-\varepsilon_i)\cdot \rho_{\max}.
\end{aligned} \tag{4}$$

The corresponding changes of the triangular FD that may occur at each section and traffic direction are illustrated in Fig. 3. More specifically, when the value of control input is 0.5, i.e., the flow capacities of the two directions are equal, their FDs are "nominal" (blue line with $(.)^N$ parameters); when the control input is different than 0.5, we have two FDs: the extended one (green line with $(.)^E$ parameters) applies to the direction that is assigned higher width, and the reduced, complementary FD (orange line with $(.)^R$ parameters) applies to the other direction that is assigned less flow capacity. Based on (4), all FD parameters of a section change, whenever it is decided to change the corresponding sharing factor in real time, i.e. at discrete times $k_c=1,2,.....$

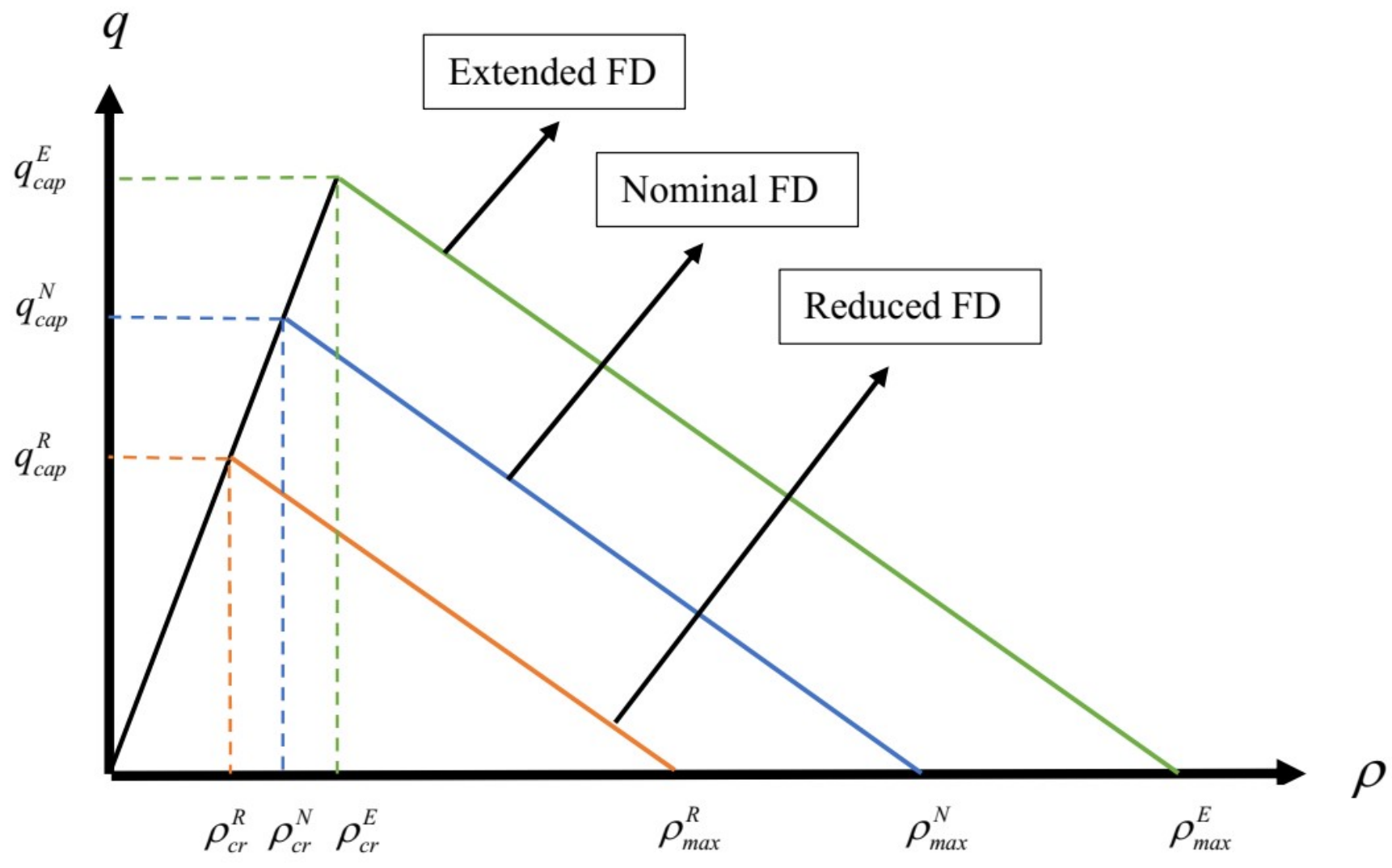

*Figure 3. The triangular fundamental diagram with flexible internal boundary.*

By their nature, the sharing factors take values $\varepsilon_i \in (0,1)$. However, for the internal boundary control problem, we would like to disallow the utter closure of either direction; hence, the assigned road width in either direction should never be smaller than the widest vehicles driving on the road. This requirement gives rise to stricter constraints for the sharing factors as follows

$$0 < \varepsilon_{i,\min} \le \varepsilon_i \le \varepsilon_{i,\max} < 1, \tag{5}$$

where $\varepsilon_{i,\min} \cdot w$ and $(1-\varepsilon_{i,\max}) \cdot w$ are the minimum admissible widths to be assigned to directions $a$ and $b$, respectively. If the two minimum widths are equal, then we have $\varepsilon_{i,\min} + \varepsilon_{i,\max} = 1$. The minimum admissible width does not necessarily define a lane. The highway holds vehicles of different dimensions (trucks, small and big cars, vans, busses, motorcycles), the widest of which should be able to drive on a road with width equal to the minimum admissible width.

Another restriction to be applied to the sharing factors concerns the time-delay needed to evacuate traffic on the direction that receives a restricted width, compared with the previous control time step. As discussed earlier, this time-delay may be considerable in the case of long reversible lanes in lane-based traffic with physical reversible lane separation (Frejo et al., 2015). In contrast, this time-delay is much smaller in lane-free CAV traffic with moderate changes of the sharing factors that are applied to short sections; but needs nevertheless to be considered. Clearly, the time-delay should apply only to the traffic direction that is being widened, compared to the previous control interval; while the direction that is restricted should promptly apply the smaller width, so that CAVs therein may move out of the reduced-width zone. Assume that the required time-delay is smaller than or equal to the control time interval $T_c$; then, the time-delay requirement is automatically fulfilled for each section $i$, if the sharing factors that are actually applied to the two directions, i.e. $\varepsilon_i^a$ and $\varepsilon_i^b$, respectively, are calculated as follows

$$\varepsilon_i^a(k_c) = \min\{\varepsilon_i(k_c), \varepsilon_i(k_c - 1)\}, \tag{6}$$

$$\varepsilon_i^b(k_c) = \min\{1-\varepsilon_i(k_c), 1-\varepsilon_i(k_c - 1)\}. \tag{7}$$

These equations may be readily extended if the required time-delay is a multiple of the control time interval $T_c$.

We are now ready to present the CTM equations, considering the changing sharing factors and their constraints. We recall that we consider a highway stretch with $n$ sections, with respective lengths $L_i$. Traffic flows from section 1 to section $n$ for direction $a$; and from section $n$ to section 1 for direction $b$ (see Fig. 4 for an example). We denote $\rho_i^a$, $i = 1, 2, ...., n$, the traffic density of section $i$, direction $a$; and $\rho_i^b$, $i = 1, 2, ...., n$, the traffic density of section $i$, direction $b$. Similarly, we have the mainstream exit flows of section $i$ being denoted $q_i^a$ for direction $a$ and $q_i^b$ for direction $b$. Thus, $q_0^a$ is the feeding upstream mainstream inflow for direction $a$; and $q_{n+1}^b$ is the feeding upstream mainstream inflow for direction $b$. In addition, every section may have an on-ramp or an off-ramp at its upstream boundary. The on-ramp flow (if any) for section $i$, direction $a$, is denoted $r_i^a$; and the on-ramp flow (if any) for section $i$, direction $b$, is denoted $r_i^b$. The off-ramp flow (if any) of section $i$, direction $a$, is calculated based on known exit rates $\beta_i^a$ multiplied with the upstream-section flow, i.e. $\beta_i^a q_{i-1}^a$; and the off-ramp flow (if any) of section $i$, direction $b$, is calculated based on known exit rates $\beta_i^b$ multiplied with the upstream-section flow, i.e. $\beta_i^b q_{i+1}^b$.

We now deliver the CTM equations for each direction of the highway stretch.

*Direction $a$:*

The conservation equations for the sections of direction $a$ read:

$$\rho_i^a(k+1) = \rho_i^a(k) + \frac{T}{L_i}((1-\beta_i^a)q_{i-1}^a(k) - q_i^a(k) + r_i^a(k)), i = 1, 2, ..., n. \tag{8}$$

According to CTM, the traffic flows are obtained as the minimum of demand and supply functions, except for the last section, where we consider only the demand function, as we assume that the downstream traffic conditions are uncongested. Clearly, when writing the demand and supply functions for the case of internal boundary control, we need to consider the impact of the respective sharing factors $\varepsilon_i^a(k_c)$ on the FDs. Thus we have

$$q_i^a(k) = \min\left\{ Q_D(\rho_i^a(k), \varepsilon_i^a(k_c)), \frac{Q_S(\rho_{i+1}^a(k), \varepsilon_{i+1}^a(k_c))}{(1-\beta_{i+1}^a)} - \lambda_r r_{i+1}^a(k) \right\},\ i = 1, 2, ...., n-1,$$
$$q_n^a(k) = Q_D(\rho_n^a(k), \varepsilon_n^a(k_c)), \tag{9}$$

where the demand and supply functions are given by the following respective equations

$$Q_D(\rho_i^a(k), \varepsilon_i^a(k_c)) = \min\left\{ \varepsilon_i^a(k_c) q_{cap} + \lambda_d q_{cap} \frac{\rho_i^a(k) - \varepsilon_i^a(k_c)\rho_{cr}}{\rho_{cr} - \rho_{\max}}, v_f \rho_i^a(k) \right\},$$
$$Q_S(\rho_i^a(k), \varepsilon_i^a(k_c)) = \min\left\{ \varepsilon_i^a(k_c) q_{cap}, w_s(\varepsilon_i^a(k_c)\rho_{\max} - \rho_i^a(k)) \right\}, \tag{10}$$

while $v_f$ is the free speed (which is assumed equal for all sections for simplicity) and $w_s$ is the back-wave speed.

It is well-known that CTM does not reproduce the capacity drop, i.e. the empirical finding that, at the head of congestion, the observed flow in real traffic is reduced compared to the carriageway capacity. Capacity drop is deemed to occur in conventional traffic due to bounded and differing accelerations of different vehicles (Yuan et al., 2015; Yuan, 2016). Recently, CTM has been extended in a number of possible ways to enable the reproduction of capacity drop, see (Kontorinaki et al., 2017) for an overview and comparison. The presence of capacity drop in conventional motorway traffic is a major reason for infrastructure degradation and for the need of introducing traffic control measures to restore capacity (Papageorgiou et al., 2003; Papageorgiou et al., 2008). In contrast, in the present context of internal boundary control, the presence of capacity drop is a secondary source of amelioration of the traffic conditions, because the potential benefits achievable via opportune capacity sharing are expected to be much higher. In fact, it is unknown at the moment, if and to what extent capacity drop may occur in lane-free CAV traffic. To be able to investigate the impact of possible capacity drop, we have incorporated in the above equations the option of introducing capacity drop according to (Kontorinaki et al., 2017). More specifically, this option is enabled via the parameters $\lambda_d$ and $\lambda_r$ in the above equations. If these parameters are set $\lambda_r = 1$ and $\lambda_d = 0$, no capacity drop is reproduced, as typical for CTM; if these values are set between 0 and 1, a corresponding level of capacity drop is produced by the model.

*Direction $b$ :*

The equations for direction $b$ are analogous to those of direction $a$, with few necessary index modifications. Section numbers in direction $b$ are descending, hence we have

$$\rho_i^b(k+1) = \rho_i^b(k) + \frac{T}{L_i}((1-\beta_i^b)q_{i+1}^b(k) - q_i^b(k) + r_i^b(k)), i = 1,2,...,n, \tag{11}$$

and the flows are given by

$$\begin{aligned} q_i^b(k) &= \min\left\{Q_D(\rho_i^b(k), \varepsilon_i^b(k_c)), \frac{Q_S(\rho_{i-1}^b(k), \varepsilon_{i-1}^b(k_c))}{(1-\beta_{i-1}^b)} - \lambda_r r_{i-1}^b(k)\right\}, \ i = 2,3,....,n, \\ q_1^b(k) &= Q_D(\rho_1^b(k), \varepsilon_1^b(k_c)), \end{aligned} \tag{12}$$

where

$$\begin{aligned} Q_D(\rho_i^b(k), \varepsilon_i^b(k_c)) &= \min\left\{\varepsilon_i^b(k_c)q_{cap} + \lambda_d q_{cap}\frac{\rho_i^b(k) - \varepsilon_i^b(k_c)\rho_{cr}}{\rho_{cr} - \rho_{\max}}, v_f\rho_i^b(k)\right\}, \\ Q_S(\rho_i^b(k), \varepsilon_i^b(k_c)) &= \min\left\{\varepsilon_i^b(k_c)q_{cap}, w_s(\varepsilon_i^b(k_c)\rho_{\max} - \rho_i^b(k))\right\}. \end{aligned} \tag{13}$$

### 2.2.3. CTM-based and further linear inequality constraints

The conservation equations (8) and (11) are linear, but, due to the presence of the min-operator in (9), (10), (12) and (13), the CTM flow equations presented in the previous section are nonlinear. As proposed by Papageorgiou (1995) and practiced in most previous utilizations of CTM for optimal control (Ziliaskopoulos, 2000; Gomes and Horowitz, 2006; Roncoli et al., 2015), such nonlinearities may be transformed to linear inequalities by requesting the left-hand side of the equation, where the min-operator appears, to be smaller than or equal to each of the terms included in the min-operator. For equations (9), (10) of direction $a$, this yields the following four inequalities

$$q_i^a(k) \le v_f\rho_i^a(k), \quad i = 1,2,....,n, \tag{14}$$

$$q_i^a(k) \le \varepsilon_i^a(k_c)q_{cap} + \lambda_d q_{cap}\frac{\rho_i^a(k) - \varepsilon_i^a(k_c)\rho_{cr}(k_c)}{\rho_{cr} - \rho_{\max}}, \quad i = 1,2,....,n, \tag{15}$$

$$q_i^a(k) \le \frac{w_s}{(1-\beta_{i+1}^a)}(\varepsilon_{i+1}^a(k_c)\rho_{\max} - \rho_{i+1}^a(k)) - \lambda_r r_{i+1}^a(k), \quad i = 1,2,....,n-1, \tag{16}$$

$$q_i^a(k) \le \frac{\varepsilon_{i+1}^a(k_c) q_{cap}^{total}}{(1-\beta_{i+1}^a)} - \lambda_r r_{i+1}^a(k), \quad i = 1, 2, ...., n-1. \tag{17}$$

Similarly, we obtain from (6) the following two linear inequalities

$$\varepsilon_i^a(k_c) \le \varepsilon_i(k_c), \quad i = 1, 2, ...., n, \tag{18}$$

$$\varepsilon_i^a(k_c) \le \varepsilon_i(k_c - 1), \quad i = 1, 2, ...., n. \tag{19}$$

For direction $b$, we obtain

$$q_i^b(k) \le v_f \rho_i^b(k), \quad i = 1, 2, ...., n, \tag{20}$$

$$q_i^b(k) \le \varepsilon_i^b(k_c) q_{cap} + \lambda_d q_{cap} \frac{\rho_i^b(k) - \varepsilon_i^b(k_c)\rho_{cr}(k_c)}{\rho_{cr} - \rho_{\max}}, \quad i = 1, 2, ...., n, \tag{21}$$

$$q_i^b(k) \le \frac{w_s}{(1-\beta_{i-1}^b)} (\varepsilon_{i-1}^b(k_c)\rho_{\max} - \rho_{i-1}^b(k)) - \lambda r_{i-1}^b(k), \quad i = 2, 3, ...., n, \tag{22}$$

$$q_i^b(k) \le \frac{\varepsilon_{i-1}^b q_{cap}}{(1-\beta_{i-1}^b)} - \lambda r_{i-1}^b(k), \quad i = 2, 3, ...., n, \tag{23}$$

$$\varepsilon_i^b(k_c) \le 1 - \varepsilon_i(k_c), \quad i = 1, 2, ...., n, \tag{24}$$

$$\varepsilon_i^b(k_c) \le 1 - \varepsilon_i(k_c - 1), \quad i = 1, 2, ...., n. \tag{25}$$

In addition, the constraints (5) must be considered to appropriately limit the sharing factors.

In summary, inequalities (14) and (15) for direction $a$, and (20) and (21) for direction $b$ represent the demand part of the FD; while inequalities (16) and (17) for direction $a$, and (22) and (23) for direction $b$ represent the supply part of the FD.

It must be emphasized, however, that the min-operator is equivalent to the corresponding set of linear inequalities only if one of the inequality constraints is actually activated. If none of the inequalities is activated, then the corresponding flow takes a lower value than the one prescribed by CTM, and this occurrence, known as the flow holding-back effect, has been addressed in different ways in the literature (Doan and Ukkusuri, 2012). Since, in presence of holding-back, CTM is accordingly distorted, we consider, in the optimal control problem formulation, non-negativity and upper-bound constraints for all flow and density variables to ensure that non-physical values are excluded from the solution.

Holding-back may occur in the solution of the QP-problem if it is beneficial for the cost function, in particular for minimization of the Total Time Spent (TTS). Holding-back may indeed be beneficial in the present context if there is congestion in the QP-problem solution and if capacity drop is activated in the model. In fact, if no congestion is present, holding back traffic can only produce unnecessary delays and TTS increase. In presence of congestion, but without capacity drop, the incentive to hold back traffic is limited to special circumstances. In contrast, in presence of congestion and capacity drop, holding back traffic may be beneficial because it may mitigate the capacity drop and restore capacity flow at the bottleneck location; something that is indeed a major motivation for traffic control in conventional traffic by use of various active holding-back control measures, such as VSL (variable speed limits) and ramp metering (Papageorgiou et al., 2008).

The presence of holding-back in the QP-problem solution could be interpreted as an opportunity to apply, in addition to internal boundary control, also speed control, something that is not difficult in CAV traffic, see also (Han et al. 2017). As the primary focus of this article is to elaborate on the features of internal boundary control, we will adopt the following policy in the investigations of Section 3:

- On one hand, we will report on the direct outcome of the QP-problem solution. If this solution contains no holding-back, then it is equivalent to the CTM simulation outcome;
- On the other hand, if flow holding back is observed in the QP-problem solution, then we will use the delivered sharing factor trajectories to feed the CTM equations and obtain the corresponding traffic states without holding-back. Thus, we will have two TTS values: one stemming from the QP-problem solution (with possible holding-back that could be interpreted as the result of additionally applying speed control); and a second obtained from CTM equations fed with the QP-optimal ε-trajectories. The deviation between these two values will indicate the extent of holding-back; or, in other words, the additional benefit that could be obtained by applying speed control on the top of internal boundary control.

Finally, the following inequalities are introduced to guarantee that a reduction of the width for any section and direction will not result in a density value that exceeds the jam density:

$$\rho_i^a(k+1) \le \varepsilon_i^a\left(k_c\right)\rho_{\max}, \quad i=1,2,...,n, \tag{26}$$

$$\rho_i^b(k+1) \le \varepsilon_i^b\left(k_c\right)\rho_{\max}, \quad i=1,2,...,n. \tag{27}$$

### 2.2.4. Objective function and QP problem formulation

The objective function to be minimized is defined as follows:

$$\begin{aligned} J = {} & T\sum_{k=1}^{K}\sum_{i=1}^{n}\left(L_i\rho_i^a(k)+L_i\rho_i^b(k)\right) - w_1\sum_{k_c=0}^{K_c-1}\sum_{i=1}^{n}\left(\varepsilon_i^a(k_c)+\varepsilon_i^b(k_c)\right) \\ & + w_2\sum_{k_c=1}^{K_c-1}\sum_{i=1}^{n}\left(\varepsilon_i(k_c)-\varepsilon_i(k_c-1)\right)^2 + w_3\sum_{k_c=0}^{K_c-1}\sum_{i=2}^{n}\left(\varepsilon_i(k_c)-\varepsilon_{i-1}(k_c)\right)^2 \\ & + w_4\sum_{k_c=0}^{K_c-1}\sum_{i=1}^{n}\left(\frac{\varepsilon_i(k_c)^2}{d_i^a(k_c)}+\frac{(1-\varepsilon_i(k_c))^2}{d_i^b(k_c)}\right). \end{aligned} \tag{28}$$

The proposed cost function extends over a time horizon of $K$ model time steps or $K_c$ control time steps, where $K_c = K \cdot T/T_c$ ; it includes five terms, the first two being linear and the rest of them quadratic. We will now comment on each of these terms.

The first term represents the TTS. This is the most important term in the cost function as it determines the traffic efficiency resulting from the proposed control action. TTS attains its minimum value if the road is shared among the two directions in such a way that no congestion, which would give rise to according vehicle delays, is present in any section and direction during the time horizon. If, despite optimal internal boundary control, the creation of congestion is inevitable due to strong external demands, then TTS minimization leads to minimization of the incurred delays.

As mentioned earlier, equations (6) and (7) were replaced by the linear inequalities (18), (19) and (24), (25), respectively. This transformation bears the risk that the resulting $\varepsilon_i^a$, $\varepsilon_i^b$ might not activate either of the two values of the original min-operator, in which case a part of the road width would remain unexploited. To suppress this possibility, the second term of the objective function is introduced (with a sufficiently high weight $w_1$) to ensure that at least one of the inequalities (18), (19) and at least one of the inequalities (24), (25) will be activated, i.e. that one of the two terms included in each min-operator (6) and (7), respectively, will actually materialize.

The next three terms are quadratic and reflect secondary (as contrasted to the TTS minimization) operational and policy objectives. Therefore, the weights of these terms must be selected small enough, so that their impact on the TTS outcome is small (if any). The first quadratic term penalizes the variation of the control input in consecutive time steps, so that changes of the internal boundary of each section from one control time step to the next remain small. The second quadratic term penalizes the space variation of the control input from section to section, so as to suppress strong changes in the road width assigned to each direction within a short distance. The need for smooth space-time changes of the internal boundary is motivated by safety and convenience concerns. However, related implications will be addressed in more detail in the forthcoming microscopic testing of the concept.

The last quadratic term is policy related. Consider cases where appropriate sharing factors at the highway sections provide sufficient respective capacities for the two directions to utterly avoid congestion forming. As it will become clearer in Section 3, in such cases there is a range of possible $\varepsilon_i$ values that lead to minimum TTS, i.e. the TTS-minimum is not unique. Under these conditions, a policy question arising is: Which one out of those TTS-optimal $\varepsilon_i$-values is preferable? One possible answer to this question is to express a preference for $\varepsilon_i$-values that are closer to 0.5, i.e. closer to the middle of the road. For such a policy, this quadratic term should be chosen as $(\varepsilon_i - 0.5)^2$.

In contrast, the policy pursued with the last quadratic term in (28) is a different one, as it attempts to assign to the two directions respective capacity shares that balance the respective capacity reserves for each section. To this end, we use $d_i^a(k_c)$ and $d_i^b(k_c)$, $i = 1,..,n$, which are the projected demand trajectories for each section of the two respective directions, and are calculated as follows. For a given highway stretch and demand scenario, we consider the respective upstream mainstream entry flow, as well as all on-ramp and off-ramp flows over the whole time horizon in each direction; then, the projected demand trajectories are obtained by running the CTM equations, with these entering and exiting flows, assuming that the capacity is sufficiently high. In other words, the projected demands in each section and direction are obtained by propagating the external demands at free speed. Then, for given projected demands, the minimization of the last quadratic term w.r.t. $\varepsilon_i$ is readily seen to lead to

$$\frac{\varepsilon_i(k_c)}{d_i^a(k_c)} = \frac{1-\varepsilon_i(k_c)}{d_i^b(k_c)},$$

hence

$$\frac{\varepsilon_i(k_c)}{d_i^a(k_c)} q_{cap} = \frac{1-\varepsilon_i(k_c)}{d_i^b(k_c)} q_{cap},$$

which corresponds indeed to the balancing of the capacity reserves in the two traffic directions in every section. On the other hand, if congestion is unavoidable due to strong demands, the utilized weight $w_4$ is sufficiently small, so that the impact of this term on TTS-minimization is marginal.

We are now ready to formulate the general convex QP problem for optimal internal boundary control for any highway stretch with known external (mainstream and on-ramp) demand trajectories over the considered time horizon as follows:

$$\min J = \mathbf{C}^T \mathbf{x} + \frac{1}{2} \mathbf{x}^T \mathbf{H}\, \mathbf{x} \tag{29}$$

subject to

$$\mathbf{A}_i \mathbf{x} \le \mathbf{b}_i \tag{30}$$

$$\mathbf{A}_e \mathbf{x} = \mathbf{b}_e \tag{31}$$

$$\mathbf{b}_{lb} \le \mathbf{x} \le \mathbf{b}_{ub} \tag{32}$$

where $\mathbf{x}$ is the decision vector including all the states (densities), flows and control variables (sharing factors) at all times. The two terms in the cost function (29) represent the linear and quadratic terms of equation (28), respectively. The linear inequality (30) derives from all inequalities mentioned in Section 2.2.2; whereas the linear equality (31) represents the conservation equations (8) and (11). Finally, (32) provides upper and lower bounds for the decision variables, including (5) for the sharing factors.

# 3. Case studies

## 3.1. Introduction of case studies

Two scenarios are considered in order to investigate the performance of the proposed method:

- *Uncongested scenario*, where congestion is created in one or both directions without internal boundary control (no-control case); but the congestion can be utterly avoided with activation of internal boundary control. Such situations are likely to constitute the majority of real congestion cases on highways and arterials.
- *Congested scenario*, where no-control congestion can be mitigated with internal boundary control, but cannot be utterly suppressed due to strong bi-directional external demands.

The considered highway stretch is depicted in Fig. 4. It has a length of 3 km and is subdivided in 6 sections of 0.5 km each. In direction $a$, there is one off-ramp in section 2 and one on-ramp in section 5. In direction $b$, there is one off-ramp in section 4 and one on-ramp in section 3. The exit rates of the off-ramps are both equal to 0.1. The modelling time step is $T = 10\,s$, and the control time step $T_c = 60\,s$. The parameter values used to enable capacity drop, when this possibility is mentioned to be activated in the scenarios, are $\lambda_d = 0.4$ and $\lambda_r = 0.7$; when no capacity drop is activated in the model, these parameters are set equal to 1. The considered time horizon is 1 h, hence $K = 360$ and $K_c = 60$. The CTM parameters are $v_f = 100\,\text{km/h}$ and $w_s = 12\,\text{km/h}$; while the total cross-road capacity to be shared among the two directions is $q_{cap} = 12{,}000\,\text{veh/h}$. The upper and lower bounds for the sharing factors, so as to avoid utter blocking of any of the two directions, are equal for all sections and are given the values $\varepsilon_{\min} = 0.16$ and $\varepsilon_{\max} = 0.84$. For both scenarios, the initial density values are $\rho_i^a(0) = [5,5,5,5,18.5,29.4]\text{veh/km}$, $\rho_i^b(0) = [14.4,14.4,14,5,5,5]\text{veh/km}$.

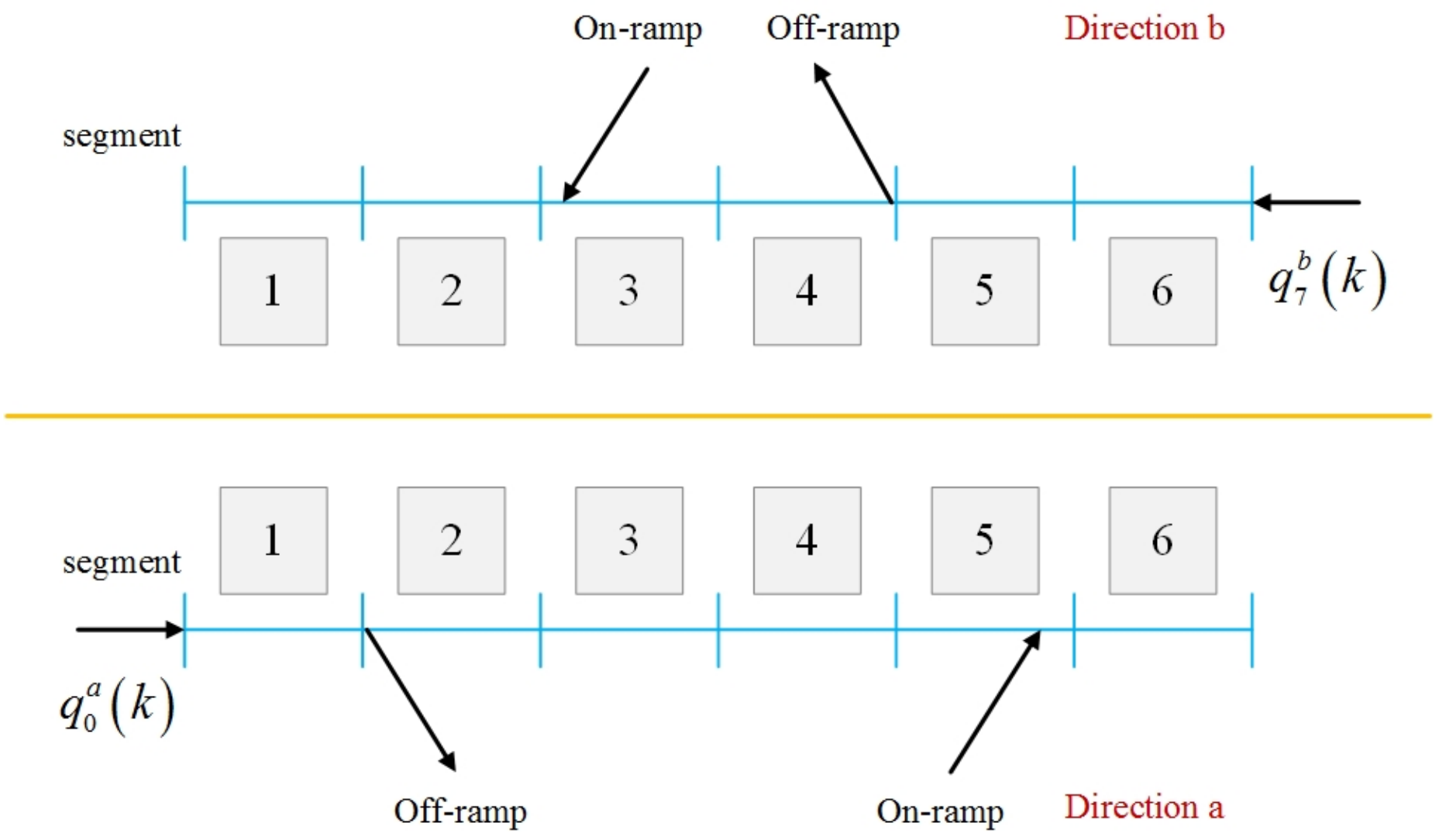

*Figure 4. The considered highway stretch*

For each scenario, the simulation results for the no-control case will be presented first, followed by the results obtained with optimal internal boundary control. The same weight parameters were used in the cost function of the QP-problem for the respective control cases of the two scenarios, namely $w_1 = 10^{-1}, w_2 = 10^{-4}, w_3 = 10^{-5}, w_4 = 10^{-3}$. These values were obtained with manual fine-tuning on the grounds outlined in Section 2.2.3, taking into account the magnitude of each related term. More specifically:

- The weight $w_1$ was selected sufficiently high to ensure that at least one of the inequalities (18), (19) and at least one of the inequalities (24), (25) is always activated, i.e. that one of the two terms included in each respective min-operator (6) and (7) actually applies.
- The weights of the three quadratic terms were initially set to zero, in order to obtain the minimum achievable TTS value. Eventually, the three weights $w_2$, $w_3$, $w_4$ were gradually increased to the above-mentioned values, such that the corresponding smoothness and balancing sub-objectives are achieved at a sufficient level; while, at the same time, TTS increases only marginally (less than 1%) compared to its minimum achievable value.

The results were found to be little sensitive around this choice; indeed, differing control behavior, if desired, may call for corresponding weight changes by one or more orders of magnitude.

The QP algorithm employed was interior point convex in MATLAB 2018 and was executed on a personal computer with processor Intel(R) Core (TM) i5-7500 CPU @ 3.4GHz and RAM of 8 GB. For an optimization horizon of 360 model time-steps (1 hour) the computational times were around 25 s.

### 3.2. Uncongested scenario

#### 3.2.1. Scenario description

The demand flows for this scenario are displayed in Fig. 5a for both directions. It may be seen that the two directions feature respective peaks in their upstream mainstream demands that are slightly overlapping. In addition, the on-ramp demands are constant, with the on-ramp demand in direction $a$ being higher than in direction $b$.

The demand and supply situation for this scenario can be analyzed based on Fig. 6. The figure displays, for each highway section, a flow-versus-time window, where time extends over the time-horizon of the scenario. The height of each window equals the total road capacity $q_{cap}$ that must be shared among the two directions at each section. The lower displayed curve (blue) is the projected demand $d_i^a(k_c)$ in direction $a$ for each section; and the upper curve (red), displayed upside-down, starting from the upper window edge, is the projected demand $d_i^b(k_c)$ in direction $b$. The fact that these two curves do not intersect at any section, indicates that flexible sharing may be applied so as to avoid any congestion forming in either direction. In fact, the finite distance among the two curves at any time for all sections indicates that there is a whole range of possible sharing factor trajectories that lead to congestion avoidance, and one such trajectory is indeed displayed (green) in each diagram, specifically one that balances approximately the capacity reserves among the two directions. In addition, each diagram displays a horizontal line (black) at 6000 veh/h, which is the capacity of each direction if no internal boundary control is applied. It may be seen that this line intersects with the projected demand curves at some sections during some time periods. Specifically, the projected demand in direction $a$ exceeds the fixed capacity first at section 5 at around $k = 60$; while in direction $b$, the projected demand exceeds capacity first in section 3 at around $k = 200$. In both cases, the exceeding of capacity is due to the presence of on-ramps in the respective sections. Obviously, these sections and time periods are candidates for congestion forming in the no-control case.

### 3.2.2. No control case

Using the entering flows of the uncongested scenario in the CTM equations of Section 2.2.1 with constant internal boundary at $\varepsilon_i = 0.5$ for all sections, we obtain the simulation results of the no-control case. Figure 7 displays the corresponding spatio-temporal density evolution. More precisely, the variable displayed in Fig. 7 for each direction is the relative density, which is defined as $\tilde{\rho}^a(k) = \rho^a(k) / \rho^a_{cr}(k) = \rho^a(k) / (\varepsilon^a(k)\rho_{cr})$ for direction $a$ and $\tilde{\rho}^b(k) = \rho^b(k) / \rho^b_{cr}(k) = \rho^b(k) / (\varepsilon^b(k)\rho_{cr})$ for direction $b$. Note that density (in veh/h) by itself is not sufficient, in the internal boundary control environment, to distinguish between under-critical and congested conditions, because the critical density is also changing according to the applied control, see (4). Of course, the critical density is not changing in the no-control case, but we use already here relative densities for consistency with the control case. According to the definition, relative density values lower than 1 refer to uncongested traffic; while values higher than 1 refer to congested traffic; clearly, when the relative density equals 1, and the downstream section is uncongested, we have capacity flow at the corresponding section.

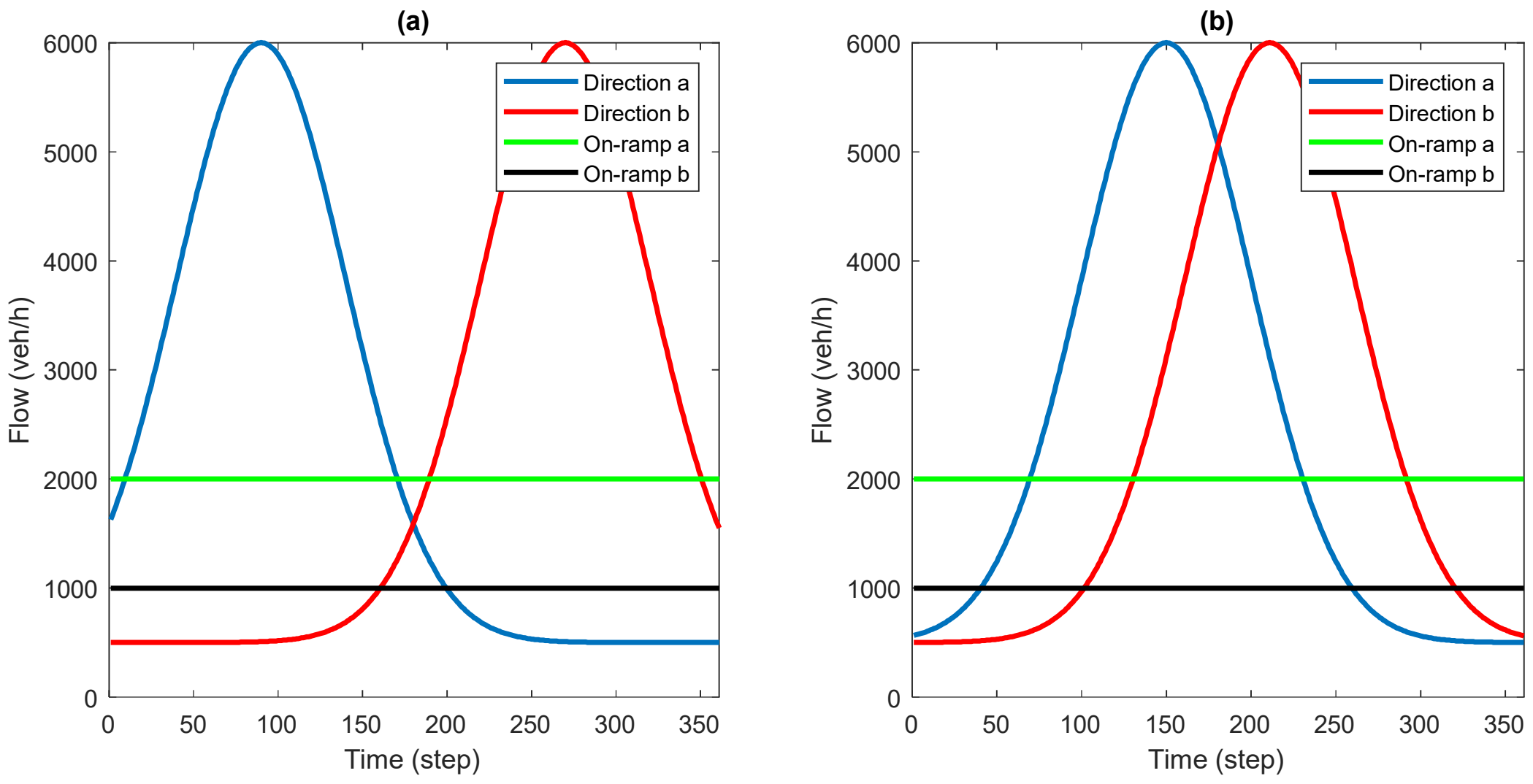

*Figure 5. Demand flows of each direction in: (a) uncongested scenario; (b) congested scenario*

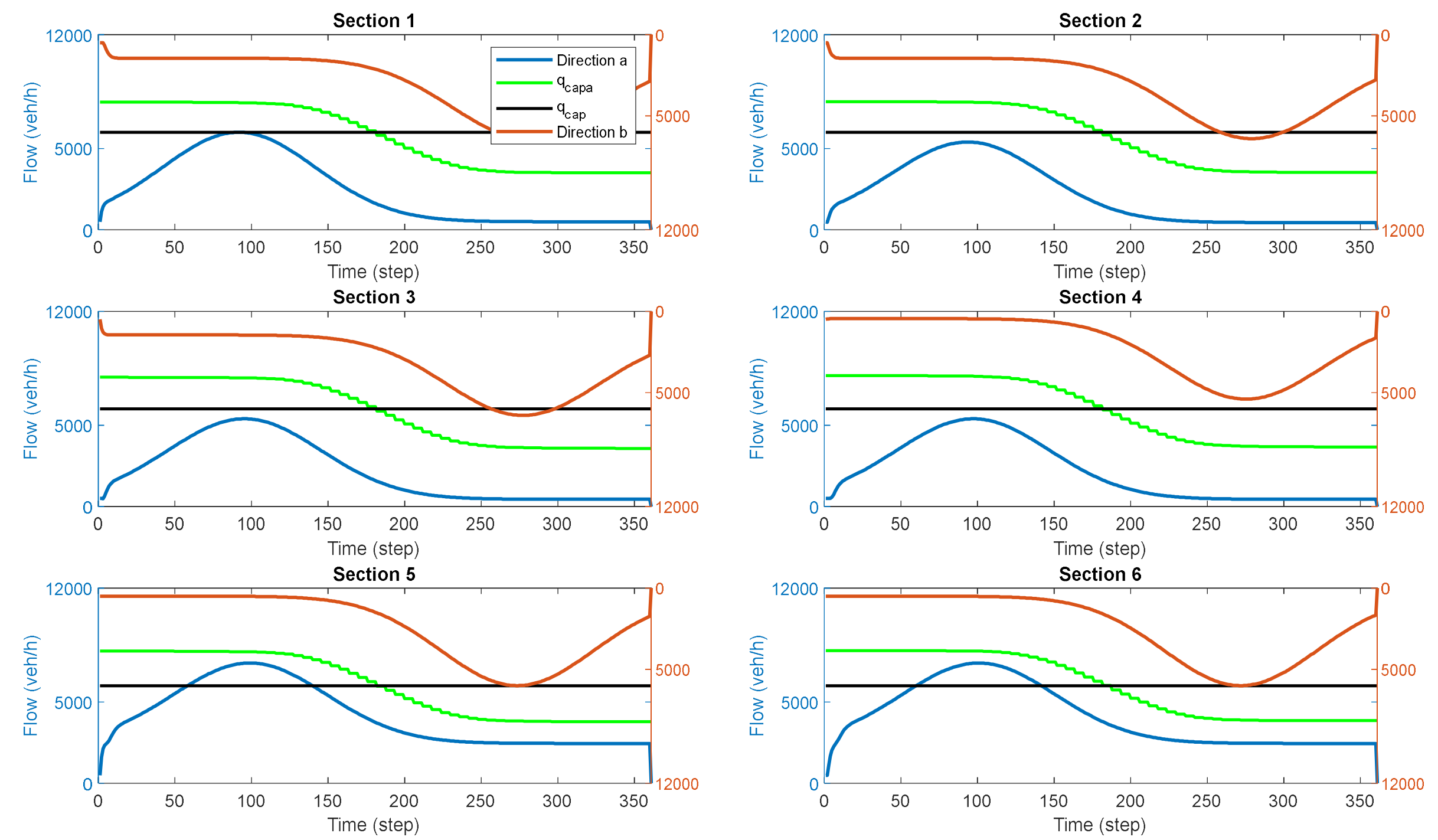

*Figure 6. Demand-supply analysis for the uncongested scenario*

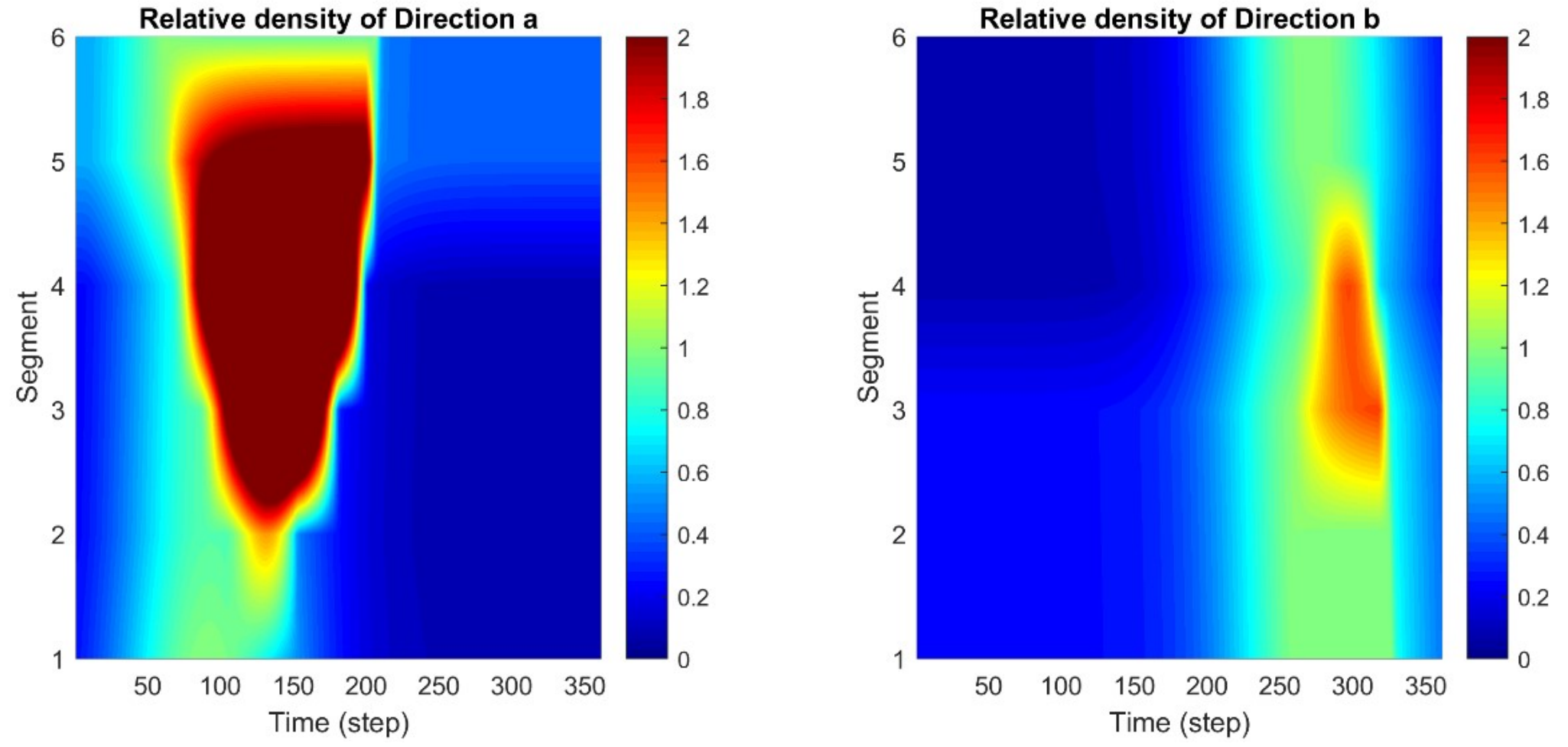

*Figure 7. Uncongested scenario: Relative density of the two directions in the no-control case*

Figure 7 shows that, as expected from the analysis with Fig. 6, heavy congestion is created in section 5 for direction $a$ due to the strong ramp inflow, in combination with the increased mainstream demand, at around $k = 60$. The congestion tail propagates backwards, reaching up to section 2, and is dissolved at around $k = 200$, thanks to the rapid decrease of the mainstream demand (Fig.5a). In direction $b$, we have also a congestion being triggered by the increasing mainstream demand, in combination with the on-ramp flow, in section 3 at around $k = 250$. Due to lower on-ramp flow, this congestion is smaller than in direction $a$; it spills back up to section 5 and dissolves at around $k = 330$.

It should be noted that the results displayed in Fig. 7 were obtained using the CTM equations with capacity drop, and the corresponding value of TTS is reported in Table 1. When the option of creating capacity drop is de-activated, then the space-time extent of the created congestions is reduced. The corresponding diagrams are omitted for space economy, but the resulting, lower TTS value is also reported in Table 1.

### 3.2.3 Control case

Based on the analysis with Fig. 6, there exist infinitely many internal-boundary trajectories that may accommodate the bi-directional demand in all sections such that the assigned capacity in each direction is never exceeded. In fact, the mentioned green curves in the diagrams of Fig. 6 reflect the obtained solution for this scenario, and it may be seen that this curve does not intersect with the projected demands in any section and direction. As a consequence, the resulting traffic conditions are expected to be under-critical everywhere, and this is indeed confirmed by the spatio-temporal evolution of the relative densities depicted in Fig. 8.

Figures 9, 10, 11 display more detailed information for this case. Specifically, each figure holds the results of two respective sections; for each section, we provide three diagrams:

- The first diagram shows the two traffic densities (in veh/km), for directions $a$ and $b$, and the corresponding two critical densities, which are changing according to the sharing factor in the section.
- The second diagram shows the two traffic flows, for directions $a$ and $b$, and the corresponding two capacities, which are changing according to the sharing factor in the section. In addition, the sum of both flows is also displayed (yellow curve).
- The third diagram shows the two sharing factors $\varepsilon^a(k_c)$ and $\varepsilon^b(k_c)$, for directions $a$ and $b$, respectively. Note that the time axis in this case displays the control time steps $k_c$.

The displayed results confirm that densities (flows) are always lower than the respective critical densities (capacities) in all sections and in both directions; hence traffic conditions are always and everywhere under-critical. In fact, the total-flow curve (for both directions) does not reach the total carriageway capacity (of 12,000 veh/h) at any time anywhere. Also, the margins of densities (flows) to the respective critical densities (capacities) are seen to be sufficiently balanced for the two directions at all sections for all time steps, taking into account that space-time smoothness of the sharing factors is also considered. In short, congestion is utterly avoided and any occurring delays in the no-control case are now utterly nullified, while all operational sub-objectives are sufficiently accounted.

Indeed, the sharing factor trajectories of the sections reveal that this excellent outcome is enabled via a smooth swapping of assigned capacity to the two directions, whereby more capacity is assigned to direction $a$ during the first half of the time horizon and vice-versa for the second half, so as to accommodate the changing respective demands and their peaks.

The reported results were obtained with activation of capacity drop, which however, has no impact, as there is not the least congestion in the QP-problem solution. For the same reason, virtually no holding-back is observed in the solution. All related TTS values are given in Table 1, indicating improvements of 29 % and 21 % over the no-control case with and without capacity drop activation, respectively.

Finally, Fig. 11a displays the space-time diagram of the control input (sharing factors) which demonstrates that it is a smooth function in space and time. Clearly, this shape may be further influenced by appropriate changes in the utilized weights of the cost criterion.

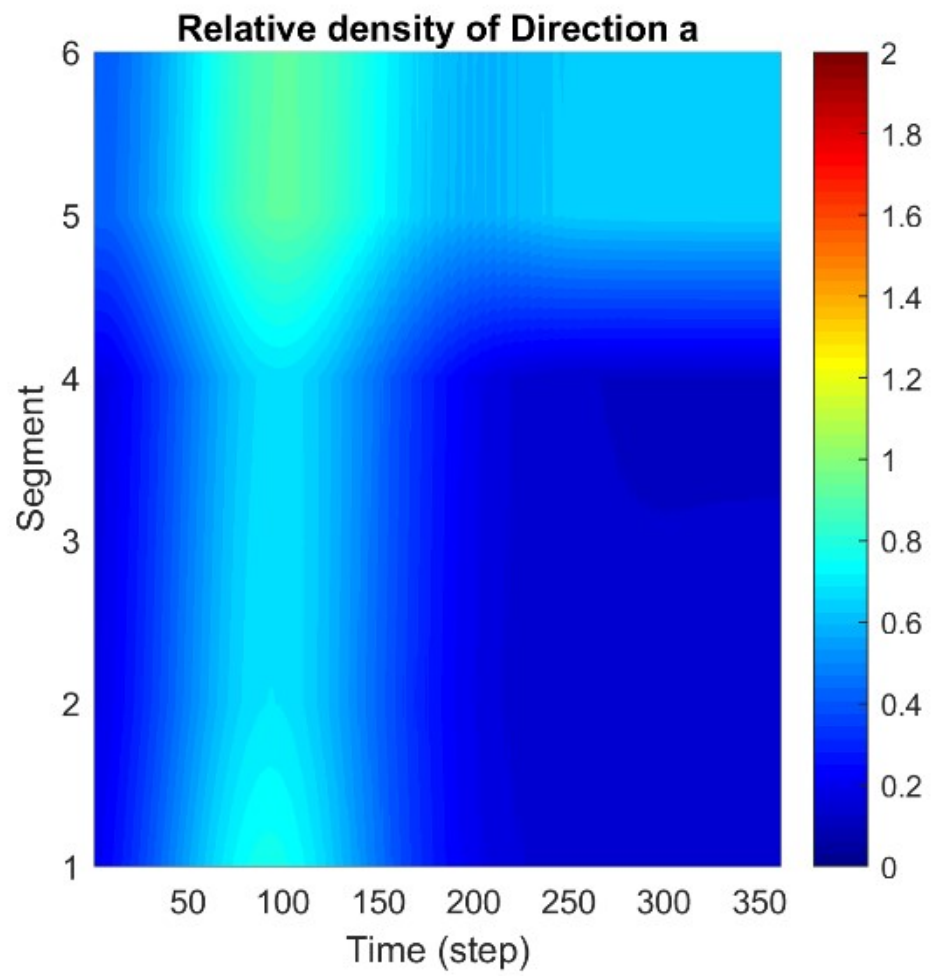

*Figure 8. Uncongested scenario: Relative density of the two directions in the control case*

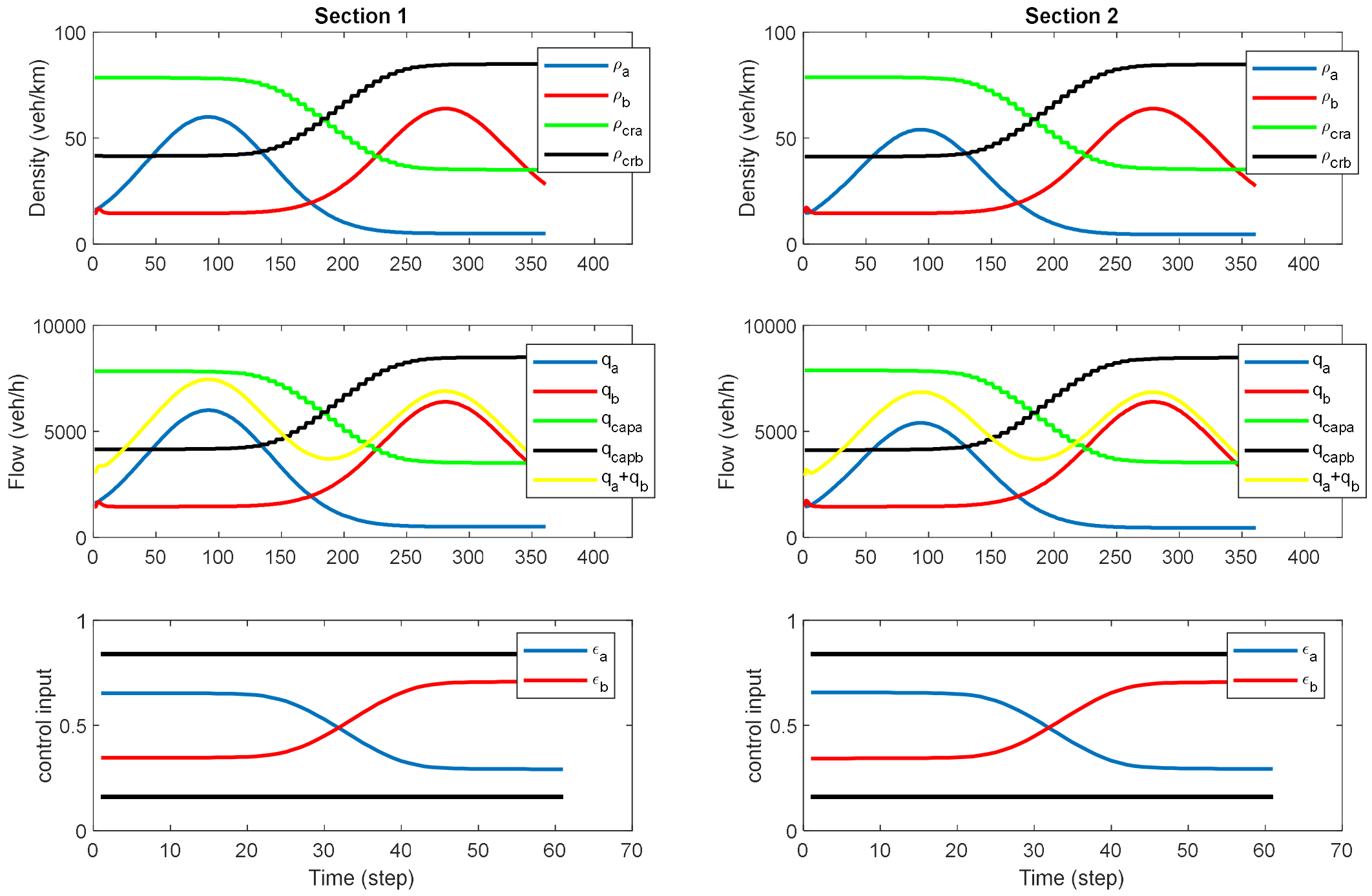

*Figure 9. Uncongested scenario: Density, flow and control trajectories in the control case (sections 1 and 2)*

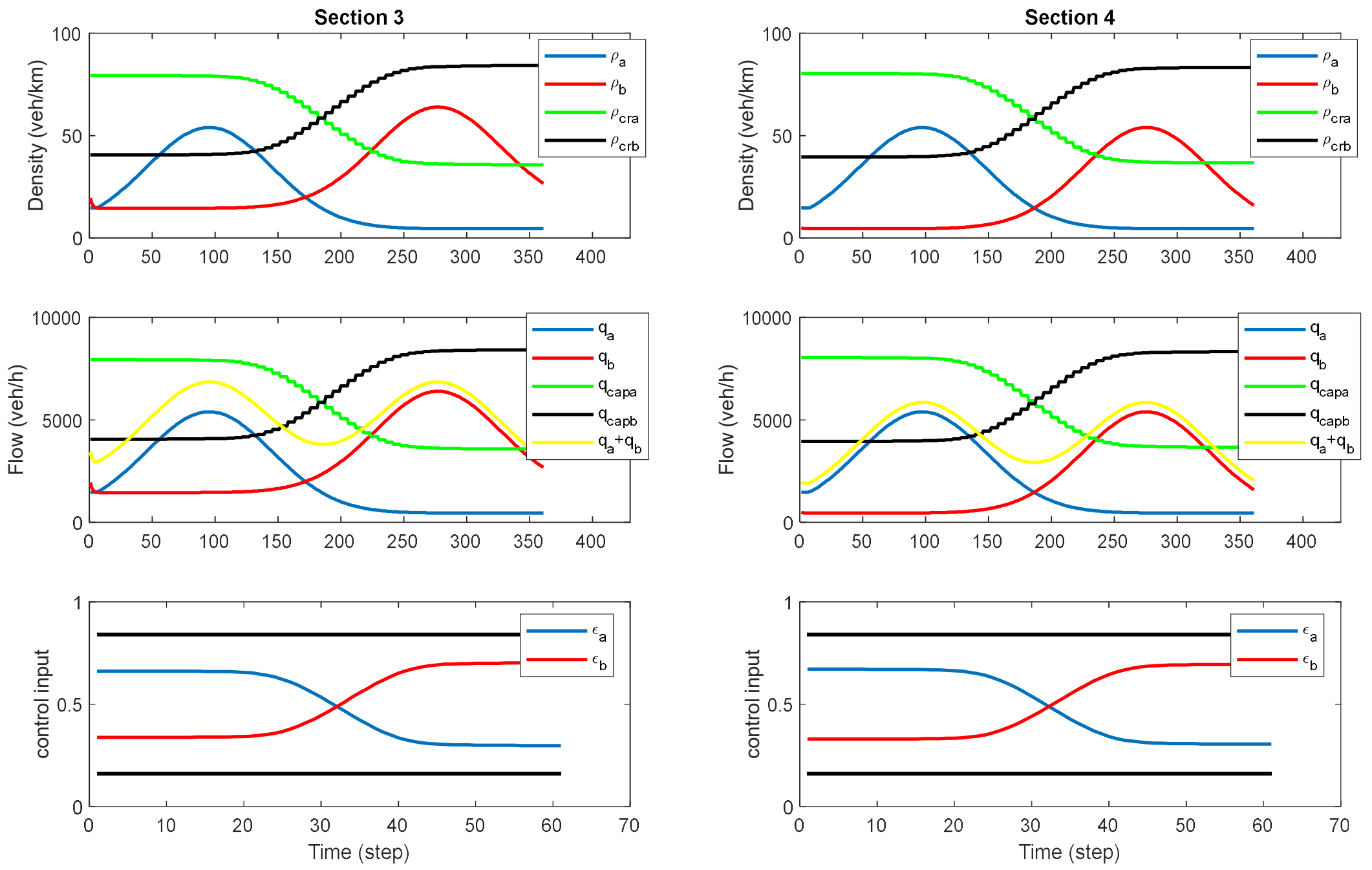

*Figure 10. Uncongested scenario: Density, flow and control trajectories in the control case (sections 3 and 4)*

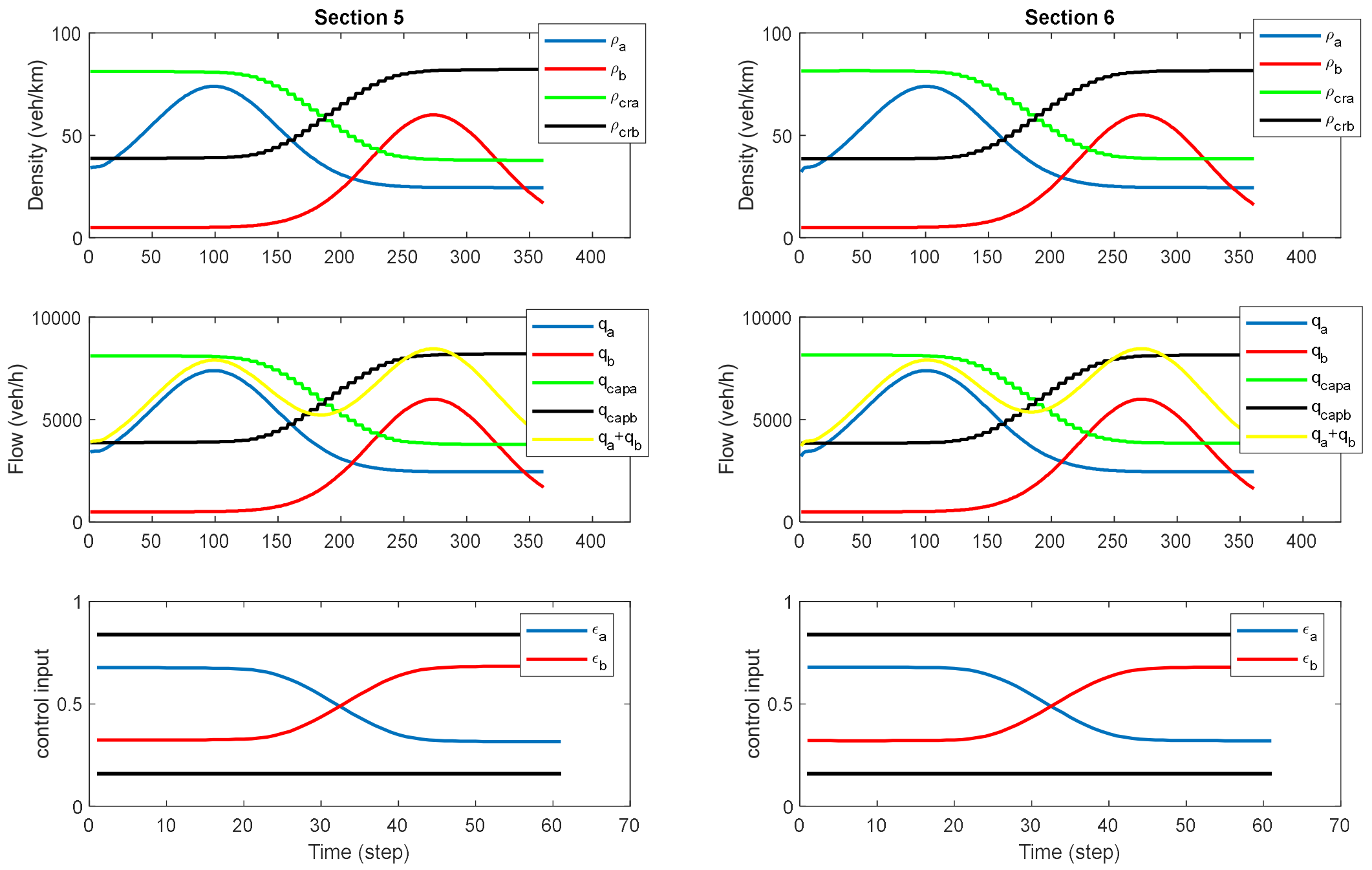

*Figure 11. Uncongested scenario: Density, flow and control trajectories in the control case (sections 5 and 6)*

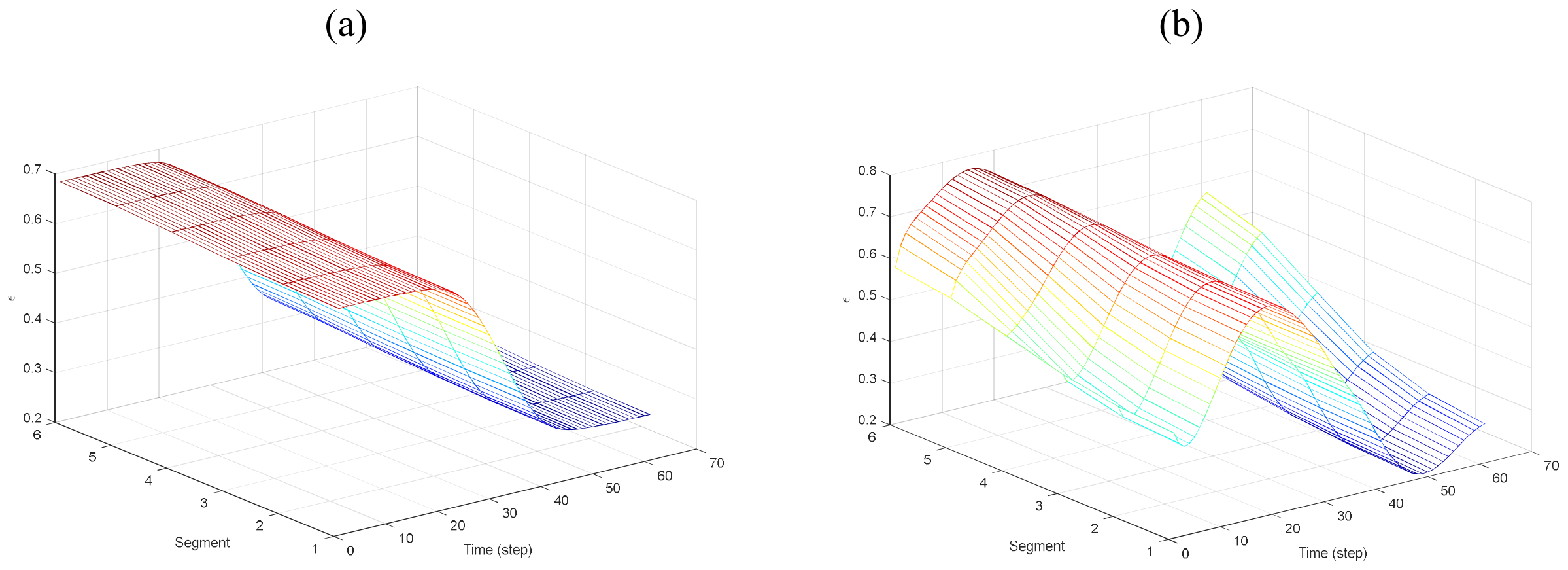

*Figure 12. 3-D space-time diagram of the control input (sharing factors) for: (a) uncongested scenario; (b) congested scenario*

## 3.3 Congested scenario

### 3.3.1 Scenario description

The demand flows for this scenario are displayed in Fig. 5b for both directions. Essentially, all external flows are similar as in the uncongested scenario, with the noticeable difference that the two mainstream demands have been moved closer to each other. This leads to a longer overlapping period with strong flows on both directions, which implies capacity problems even in presence of internal boundary control. The on-ramp demands are constant at the same level as in the uncongested scenario.

The demand and supply situation for the congested scenario can be seen in Fig. 13, which displays the same kind of information as Fig. 6. It may be seen that the no-control fixed capacity line of 6,000 veh/h intersects with the projected demand curves. Specifically, the projected demand in direction $a$ exceeds the fixed capacity first at section 5 at around $k = 120$; while in direction $b$, the projected demand exceeds capacity first in section 3 at around $k = 200$. Again, exceeding of capacity is due to the presence of on-ramps in the respective sections, and, obviously, these sections and time periods are candidates for congestion forming in the no-control case. In contrast to the uncongested scenario, we now see that the curves of the two projected demands also interfere slightly in sections 5 and 6, starting at around $k = 200$. This implies that, even with optimal boundary control, formation of congestion is unavoidable at these sections and time.

It is very interesting to emphasize here that, in contrast to conventional traffic where bottlenecks may be present in either direction independently, the application of internal boundary control implies that bottlenecks concern both traffic directions simultaneously. In other words, for a bottleneck to be present in a section, the total bi-directional projected demand must exceed the total carriageway capacity, and this obviously involves both traffic directions. Note though that a bottleneck could also occur in some cases at low demand in one traffic direction, simply because the minimum-width constraint (5) has been activated, hence no additional road width can be assigned to the opposite high-demand direction.

Figure 13 also displays a possible set of sharing factor trajectories (green) that lead to congestion avoidance anywhere except at the location and time of the identified bottleneck. These trajectories are seen to balance approximately the capacity reserves in the two directions wherever possible (i.e. at times and locations without a bottleneck).

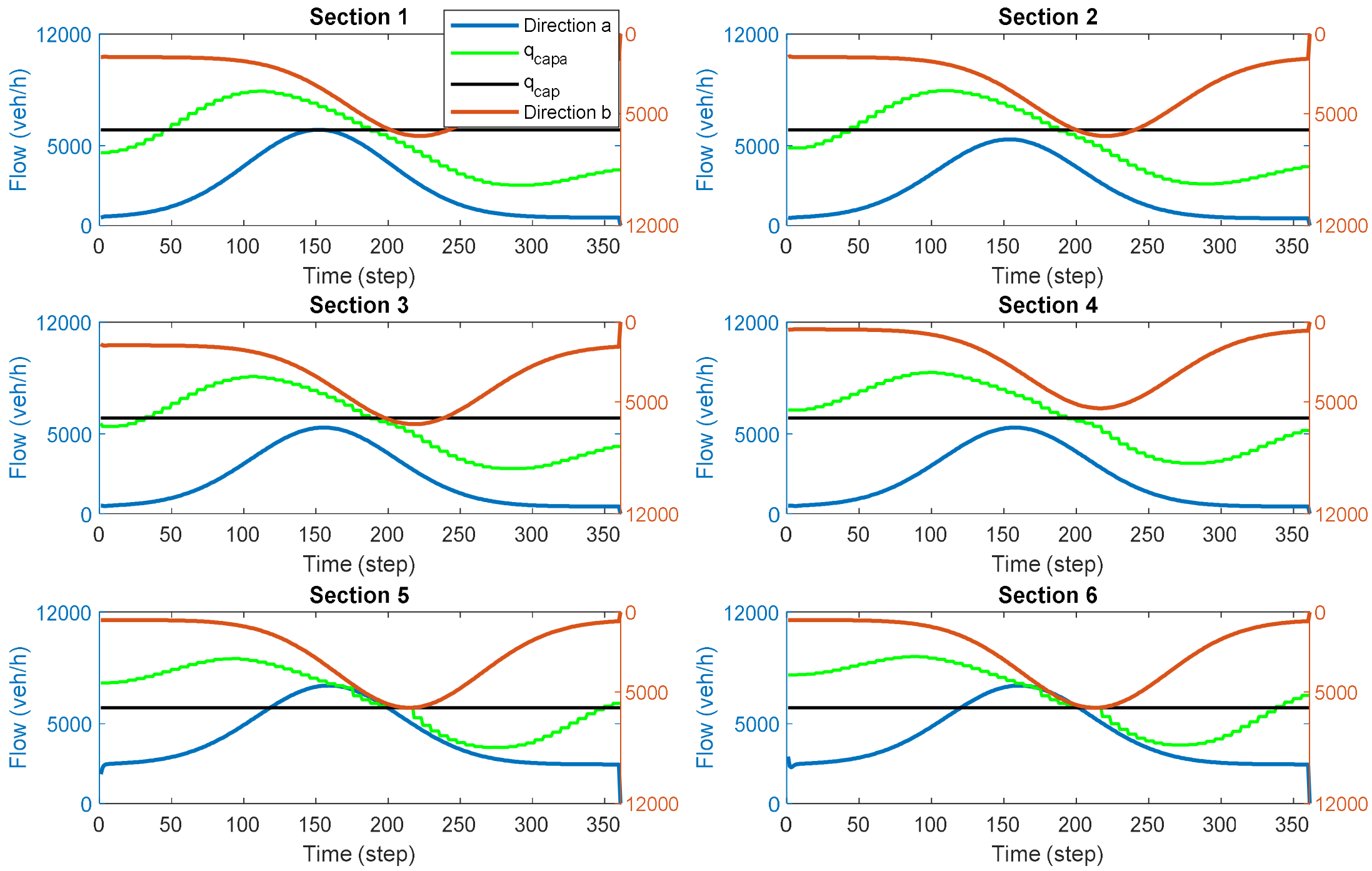

*Figure 13. Demand-supply analysis for the congested scenario,*

### 3.3.2 No control case

Using the entering flows of the congested scenario in the CTM equations of Section 2.2.1 with constant internal boundary at $\varepsilon_i = 0.5$ for all sections, we obtain the simulation results of the no-control case. Figure 14 displays the corresponding spatio-temporal relative density evolution, where, as expected from the analysis with Fig. 13, heavy congestion is created in section 5 for direction $a$ due to the strong ramp inflow, in combination with the increased mainstream demand, at around $k = 120$. The congestion tail propagates backwards, reaching up to section 2, and is dissolved at around $k = 250$, thanks to the rapid decrease of the mainstream demand (Fig.5b). In direction $b$, we have also a congestion being triggered by the increasing mainstream demand, in combination with the on-ramp flow, in section 3 at around $k = 200$. Due to lower on-ramp flow, this congestion is smaller than in direction $a$; it spills back up to section 5 and dissolves at around $k = 270$.

The results displayed in Fig. 14 were obtained using the CTM equations with capacity drop, and the corresponding value of TTS is reported in Table 1. When the option of creating capacity drop is de-activated, then the space-time extent of the created congestions is reduced. The corresponding diagrams are omitted for space economy, but the resulting, lower TTS value is also reported in Table 1. Note that, compared with the uncongested scenario, the congestion extent and TTS values are not much different, simply because the total demand per direction did not change much, while the capacity of each direction is constant in the no-control case.

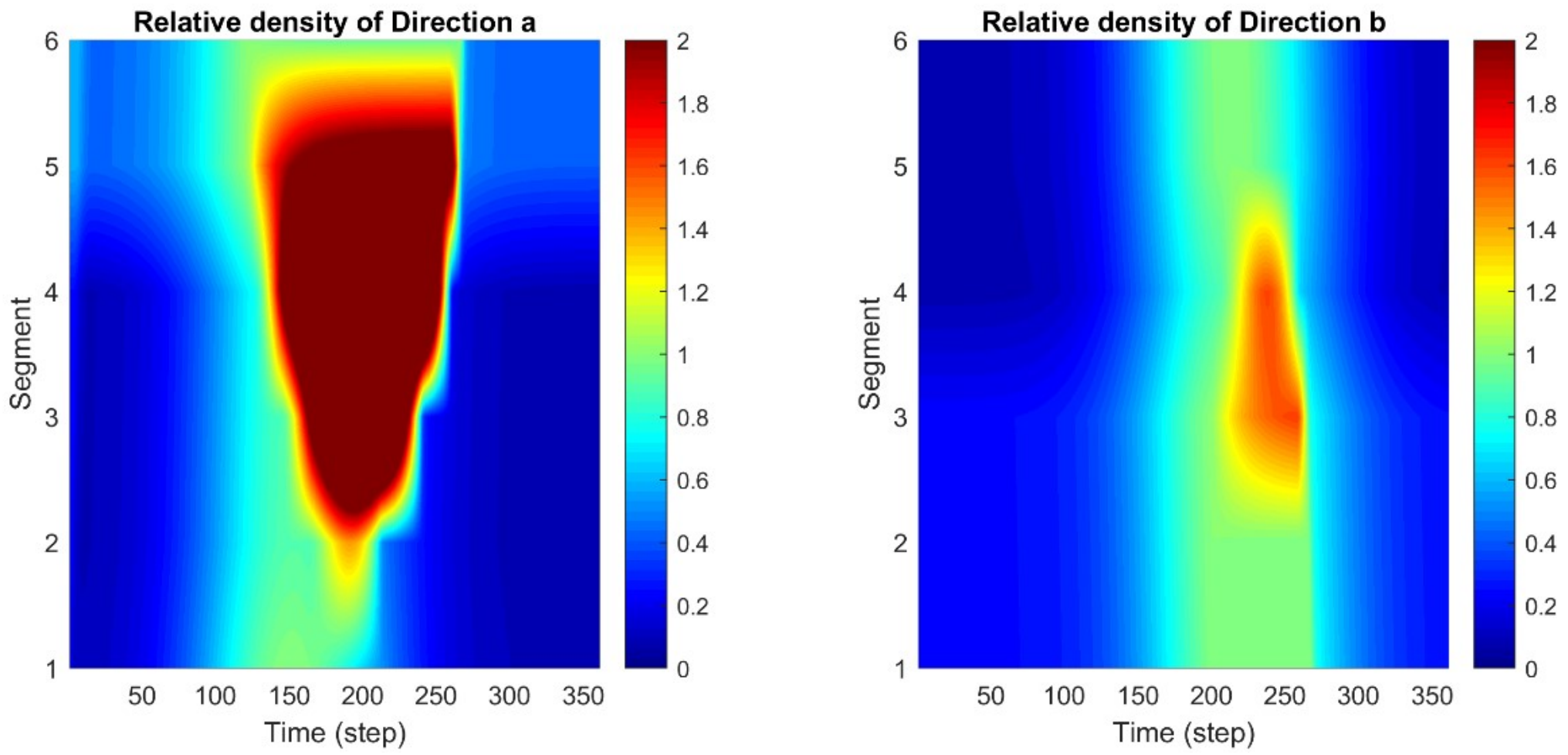

*Figure 14. Congested scenario: Relative density of the two directions in the no-control case*

### 3.3.3 Control case

Based on the analysis with Fig. 13, there exist infinitely many internal-boundary trajectories that may accommodate the projected demand so that the assigned capacity in each direction is not exceeded, with one exception, which concerns the bottleneck location and time period, as mentioned earlier. The bottleneck is common for both directions, and the sharing factors at sections 5 and 6 during the identified period should be such that the unavoidable congestion(s) affect as little as possible the resulting TTS value. In fact, the mentioned green curves in the diagrams of Fig. 13 reflect the obtained optimal solution for this scenario, and it may be seen that this curve intersects with the projected demands only at the bottleneck location and time period, while it exhibits balanced capacity margins elsewhere. As a consequence, the resulting traffic conditions are expected to be over-critical at the bottleneck location and time period, although any congestion formed there may of course propagate upstream and cover further upstream sections, depending on the amount of excess demand.

Figure 15 displays the spatio-temporal evolution of the relative densities in the control case; while Figures 16, 17, 18 display more detailed information for this case, as for the uncongested scenario. The displayed results indicate that densities (flows) are always lower than or equal to the respective critical densities (capacities) in all sections of direction $b$. In direction $a$, congestion may be observed, mainly in section 5 and to much lesser extent in section 6, starting at time $k = 170$ and lasting up to $k = 220$. It is important to emphasize that the total-flow curve (for both directions) is seen to reach and remain close to the total carriageway capacity (of 12.000 veh/h) at section 6, which is indeed the bottleneck of this scenario. Thus, full exploitation of the carriageway capacity (both directions) is indeed enabled at the bottleneck for the duration of the critical period, so as to minimize congestion and delays. Finally, the margins of densities (flows) to the critical densities (capacities) are seen to be sufficiently balanced for the two directions at any time, except for the bottleneck location and time period.

The sharing factor trajectories are slightly more varying in this scenario, compared with the uncongested scenario, so as to assign the necessary share of capacity where and when needed. Figure 12b displays the space-time diagram of the control input (sharing factors), which illustrates that it is a slightly more complex, but still smooth function in space and time.

The reported results were obtained based on the CTM equations, which were fed with the QP-optimal sharing factor trajectories. As a matter of fact, the QP problem solution contains some limited flow holding back (at section 4), which enables a slightly lower TTS value, compared with the one resulting from the CTM equations. All TTS values, namely with/without capacity drop activation and with/without holding-back, are reported in Table 1, where it may be seen that the respective differences are minor. TTS improvements over the no-control case are 28 % and 20 % with and without capacity drop activation, respectively.

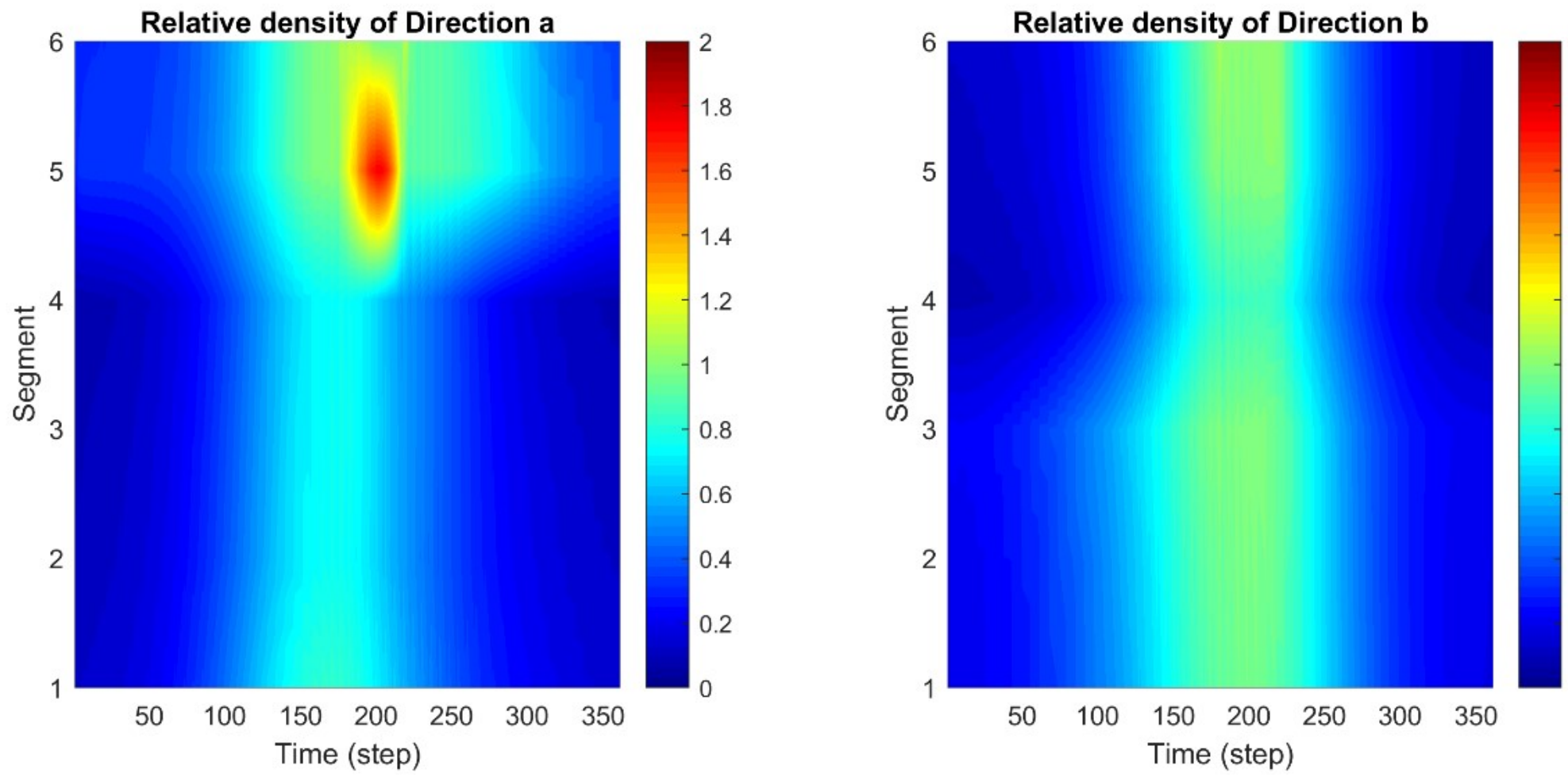

*Figure 15. Congested scenario: Relative density of the two directions in the control case*

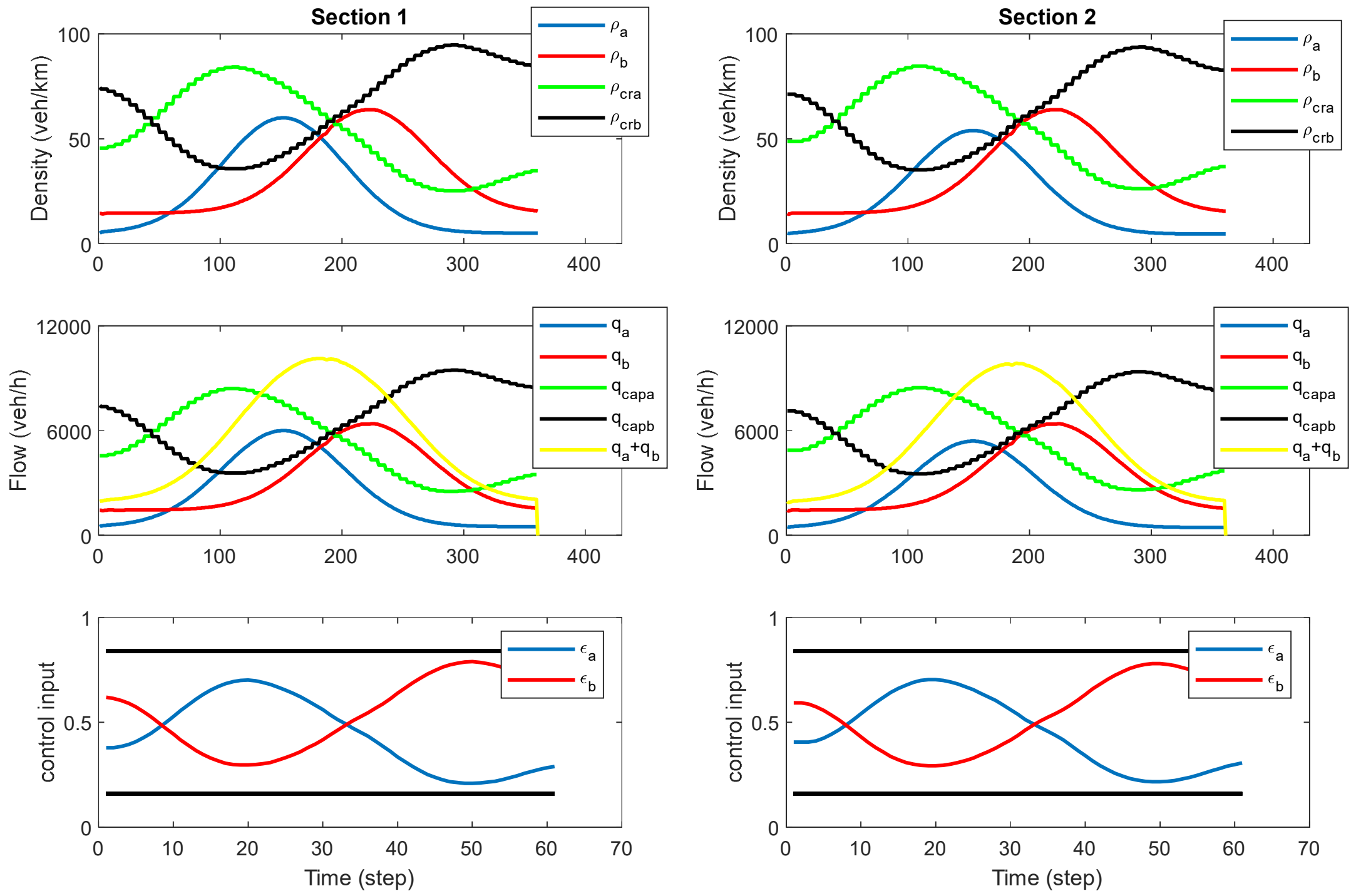

*Figure 16. Congested scenario: Density, flow and control trajectories in the control case (sections 1 and 2)*

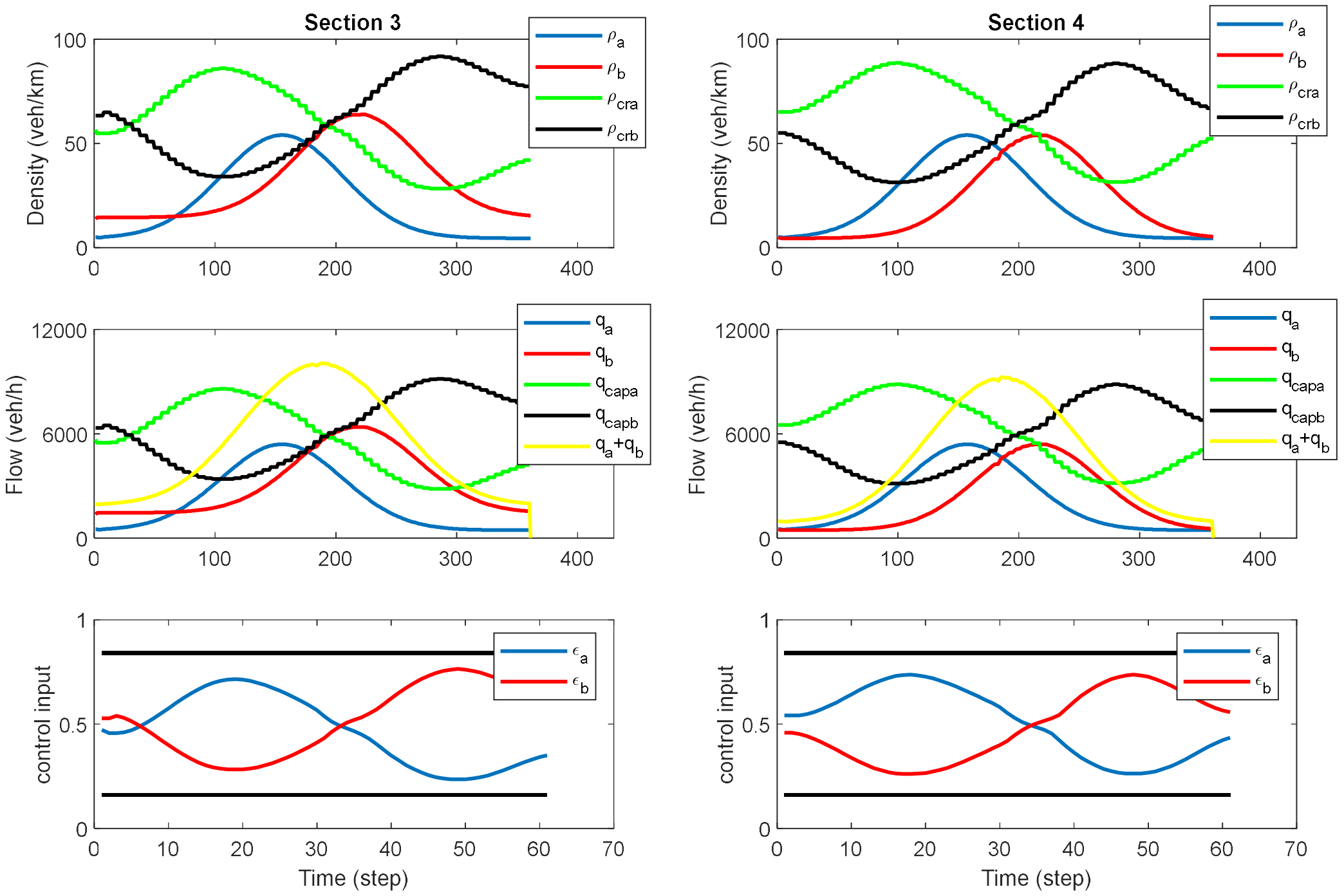

*Figure 17. Congested scenario: Density, flow and control trajectories in the control case (sections 3 and 4)*

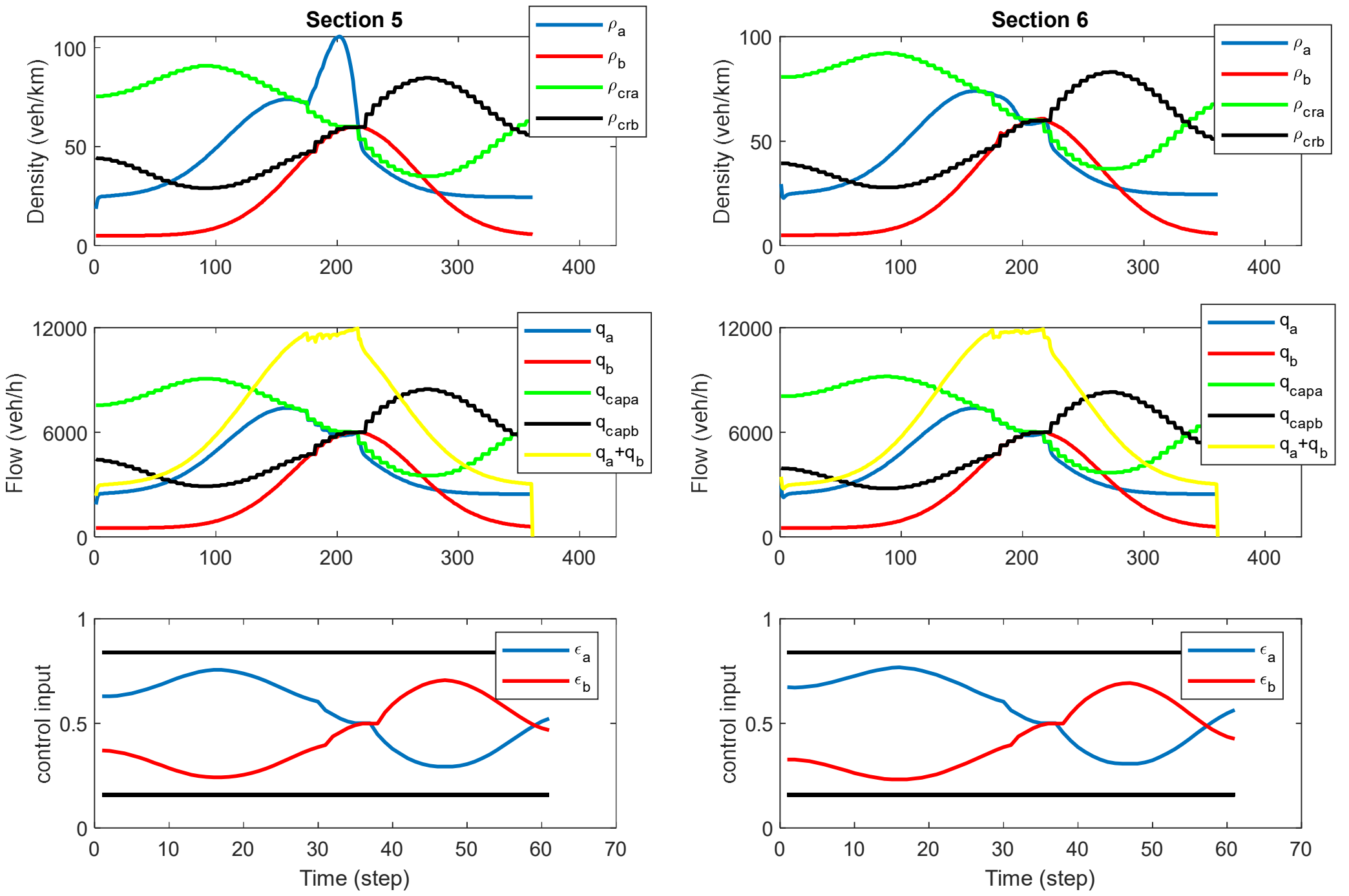

*Figure 18. Congested scenario: Density, flow and control trajectories in the control case (sections 5 and 6)*

*Table 1. The value of TTS (veh·h) and related improvement (%) over the no-control case in different scenarios*

| Scenario | No-control case | QP results | CTM simulation with QP control |
|---|---|---|---|
| Uncongested with capacity drop | 231.9 | 164.9 (-28.9 %) | 164.9 (-28.9 %) |
| Uncongested without capacity drop | 209.8 | 164.8 (-21.4 %) | 164.9 (-21.4 %) |
| Congested with capacity drop | 236.0 | 170.4 (-27.8 %) | 171.0 (-27.5 %) |
| Congested without capacity drop | 213.9 | 170.1 (-20.5 %) | 170.9 (-20.1 %) |

## 4. Conclusions

In this study, a new traffic control concept called internal boundary control has been presented, which is applicable in lane-free CAV traffic. Based on this concept, the capacity flow for each traffic direction is not constant, but can be flexibly adjusted according to the bi-directional demand and congestion conditions. For demonstration of the new concept, a quadratic programming formulation was adopted, which is fast enough for employment in a model predictive control (MPC) frame for real-time control.

Simulation case studies were designed for two scenarios that are representative for many others. Specifically, an uncongested scenario was considered first, where congestion appearing with fixed internal boundary (no-control case) can be utterly lifted with flexible boundary. This is likely the most common case in current highways and arterials, and the potential usage of the method would lead to full delay elimination in such cases. Secondly, a congested scenario was considered, where, due to strong bi-directional demand, congestion cannot be fully eliminated with optimal internal boundary control, but can be strongly mitigated with substantial benefits.

To achieve these results, the new concept necessitated the elaboration of novel definitions, notions and developments, including:

- The sharing factor, which is a novel traffic control input extending over space (sections) and time; and its inclusion into macroscopic traffic flow models, so as to reflect the corresponding traffic impact.
- The relative density, to help distinguishing under-critical from over-critical traffic conditions;
- The novel incurred notion of bottleneck, which, in contrast to conventional traffic, now refers to both traffic directions simultaneously.

Ongoing work considers a number of parallel avenues: Design of feedback control approaches for internal boundary control that operate without the need for model and external demand prediction; simulation studies with microscopic vehicles moving in a lane-free mode based on appropriate CAV movement strategies; more realistic large-scale highway infrastructure scenarios.

As mentioned earlier, Duell et al. (2016) considered a lane reversal approach, incorporated in a system-optimal route guidance problem, whereby integer variables were needed to account for segment-wise whole-lane swapping to one or the other traffic direction. The incremental internal boundary control problem for highway stretches presented in this paper provides a good basis for considering management or planning tasks at the network level, such as route guidance, dynamic traffic assignment or network design problems. The significant advantage in such an approach would be that all capacity share variables (sharing factors) would be real-valued, something that would strongly simplify the related mathematical problems and accelerate solution algorithms. In addition, internal boundary control, as a real-time traffic management measure, may be integrated with other traffic management tools, such as ramp metering and speed control, for further increase of traffic flow efficiency where possible.

**Acknowledgements**

The research leading to these results has received funding from the European Research Council under the European Union's Horizon 2020 Research and Innovation programme/ ERC Grant Agreement n. [833915], project TrafficFluid.